\documentclass[a4paper,usenatbib]{mnras}

\usepackage[T1]{fontenc}
\usepackage{ae,aecompl}

\usepackage{graphicx}	
\usepackage{amsmath}	
\usepackage{amssymb}	
\usepackage{hyperref}
\usepackage{makecell}
\usepackage{bm}
\usepackage{dcolumn}
\usepackage{ulem}

\newcommand{\hi}{H\textsc{i}}
\newcommand{\althi}{H{\normalfont\textsc{i}}}

\title[Interferometric HI IM]
{Extracting \althi\ Astrophysics from Interferometric Intensity Mapping}
\author[Z.~Chen et al.]
{Zhaoting Chen$^{1}$\thanks{E-mail: zhaoting.chen@manchester.ac.uk},
Laura Wolz$^{1}$,
Marta Spinelli$^{2,3}$,
and Steven G. Murray$^{4}$ \\
$^{1}$ Jodrell Bank Centre for Astrophysics, School of Physics and Astronomy, The University of Manchester, Manchester M13 9PL, UK \\
$^{2}$ INAF-Osservatorio Astronomico di Trieste, Via G.B. Tiepolo 11, 34143 Trieste, Italy \\
$^{3}$ IFPU - Institute for Fundamental Physics of the Universe, Via Beirut 2, 34014 Trieste, Italy\\
$^{4}$ School of Earth and Space Exploration, Arizona State University, Tempe, AZ
}

\date{Draft version: Oct. 19, 2020}

\pubyear{2020}

\begin{document}
\label{firstpage}
\pagerange{\pageref{firstpage}--\pageref{lastpage}}
\maketitle

\begin{abstract}
We present a new halo model of neutral hydrogen (\hi) calibrated to galaxy formation simulations at redshifts $z\sim0.1$ and $z\sim1.0$ that we employ to investigate the constraining power of interferometric \hi\ intensity mapping on \hi\ astrophysics. We demonstrate that constraints on the small-scale \hi\ power spectrum can break the degeneracy between the \hi\ density $\Omega_{\rm \hi}$ and the \hi\ bias $b_{\rm \hi}$. For $z\sim0.1$, we forecast that an accurate measurement of $\Omega_{\rm \hi}$ up to 6\% level precision and the large-scale \hi\ bias $b_{\rm \hi}^0$ up to 1\% level precision can be achieved using Square Kilometre Array (SKA) pathfinder data from MeerKAT and Australian SKA Pathfinder (ASKAP). We also propose a new description of the \hi\ shot noise in the halo model framework in which a scatter of the relation between the \hi\ mass of galaxies and their host halo mass is taken into account. Furthermore, given the number density of \hi\ galaxies above a certain \hi\ mass threshold, future surveys will also be able to constrain the \hi\ mass function using only the \hi\ shot noise. This will lead to constraints at the 10\% level using the standard Schechter function. This technique will potentially provide a new way of measuring the \hi\ Mass Function, independent from existing methods. We predict that the SKA will be able to further improve the low-redshift constraints by a factor of 3, as well as pioneering measurements of \hi\ astrophysics at higher redshifts.
\end{abstract}

\begin{keywords}
Cosmology; techniques: interferometric; radio lines: general
\end{keywords}

\section{Introduction}

Our understanding of the evolution of the Universe builds on the standard $\Lambda$ cold dark matter (CDM) Model. $\Lambda$CDM fittingly describes the measurements from the cosmic microwave background (CMB; e.g. \citealt{2017JCAP...06..031L, 2020A&A...641A...1P}), the expansion rate of the low-redshift Universe exhibited by Type Ia supernova (e.g. \citealt{2019MNRAS.486.2184M}), the clustering of large-scale structure (LSS; e.g. \citealt{2015MNRAS.452.1914G,2017MNRAS.465.1757G}), and other observational probes. In recent years, the constraints on the $\Lambda$CDM model have reached per cent level precision \citep{2020A&A...641A...6P, 2020arXiv200708991E}. With this increasing precision, cosmological probes capable of filling the gap between the CMB and low-redshift probes have become of great interest as we try to probe the evolution of the Universe. A relatively recently proposed way of tracing the underlying dark matter distribution, namely neutral hydrogen (\hi) intensity mapping (e.g.  \citealt{1997ApJ...475..429M, 2004MNRAS.355.1339B, 2008PhRvL.100i1303C, 2009MNRAS.397.1926W, 2010Natur.466..463C}), has shown great potential. It utilises the spin flip transition line of \hi\ at the wavelength of 21\,cm in the radio band to map the clustering of the LSS.

The general idea of intensity mapping is to map the flux density of a spectral line within relatively large pixels. It allows to survey a large area of the sky in an efficient way, suitable for the purpose of cosmological measurements. It also has the benefit of being spectroscopic in nature since the rest wavelength of the spectral line is fixed. It works for any intrinsic emission line strong enough to be detectable \citep{2017arXiv170909066K}, such as [CII] (e.g. \citealt{2012ApJ...745...49G}) and CO(1-0) (e.g. \citealt{2015ApJ...814..140K}), and most promisingly for \hi, as it abundantly traces galaxies and dark matter and has minimal line confusion.

\hi\ has historically been measured through conventional methodologies such as \hi\ galaxy surveys \citep{2005MNRAS.359L..30Z,2010ApJ...723.1359M,2015MNRAS.452.3726H} and damped Lyman$\alpha$ systems \citep{2017MNRAS.471.3428R}, yielding both global quantities, such as the overall density of \hi\ in our Universe \citep{2010ApJ...723.1359M}, and \hi-related properties of galaxies and dark matter halos \citep{2019MNRAS.486.5124O,2020MNRAS.493.1587H}. From these measurements, it has been established that \hi\ comprises roughly $10^{-4}$-$10^{-3}$ of the total energy budget in the post-ionisation Universe $z\lesssim6$ \citep{2018ApJ...864..142D}.

\hi\ intensity maps were first observed using data from the Green Bank Telescope in cross-correlation with optical galaxies at $z\sim0.8$ \citep{2013ApJ...763L..20M,10.1093/mnrasl/slt074}, as well as the 2dF Galaxy Survey in cross-correlation with intensity maps obtained by the Parkes radio telescope at $z\sim0.1$ \citep{2018MNRAS.476.3382A}. Numerous experiments targeting the 21\,cm line in the low-redshift Universe have been planned, with an increasing emphasis on intensity mapping; these include surveys using telescopes such as the Baryon Acoustic Oscillations in Neutral Gas Observations (BINGO; \citealt{2012arXiv1209.1041B}), Five-hundred-meter Aperture Spherical Radio Telescope (FAST; \citealt{2020MNRAS.493.5854H}), Canadian Hydrogen Intensity Mapping Experiment (CHIME; \citealt{2014SPIE.9145E..22B}), Tianlai \citep{2015ApJ...798...40X}, ASKAP \citep{2012MNRAS.426.3385D} and MeerKAT \citep{2016mks..confE...6J}.
The latter two experiments serve as pathfinders to the highly anticipated Square Kilometre Array (SKA), which is expected to be the most powerful telescope for mapping \hi\ throughout the post-ionization Universe \citep{2020PASA...37....7S}.

The foremost challenge for upcoming \hi\ intensity mapping experiments is the accurate extraction of the cosmological signal from the observational data. In order to control systematics and acquire an accurate  measurement of the \hi\ power spectrum, various challenges must be overcome, such as understanding properties of the instrument noise (e.g. \citealt{2018MNRAS.478.2416H}) and foreground subtraction (e.g. \citealt{2014MNRAS.441.3271W, 2016MNRAS.456.2749O, 2019MNRAS.488.5452C}). The interpretation of the \hi\ brightness temperature power spectrum itself also poses a challenge. Understanding \hi\ properties as a biased tracer of dark matter becomes a priority over cosmological inference. It is estimated that improving prior constraints on astrophysics may improve cosmological constraints by a factor of $\sim2$ \citep{ 2019MNRAS.485.4060P,2020MNRAS.495.3935P}. Thus, it is also important to explore the possibility of using intensity mapping to understand the astrophysics of \hi.

Cosmological inference requires clustering measurements primarily on large scales. Thus, a large number of \hi\ intensity mapping experiments will be operating on single-dish mode, to probe large survey volumes and to provide strong constraints on cosmological parameters (e.g. \citealt{2020PASA...37....7S}). However, interferometric observations of \hi\ intensity mapping are crucial to extract the astrophysics of \hi\ on smaller scales \citep{2020arXiv200913550P}.

Previous studies (e.g. \citealt{2015ApJ...803...21B, 2015aska.confE..19S}) often treat the \hi\ power spectrum as a simple combination of overall density and linear bias on top of the matter power spectrum, which is reasonable given the limited scale and large error of existing measurements. However, as simulations suggest (e.g. \citealt{2014JCAP...09..050V, 2020MNRAS.493.5434S, 2021ApJ...907....4W}), the simple assumption of a scale-independent linear \hi\ bias breaks down at $k\gtrsim0.3\,{\rm Mpc^{-1}}h$ \citep{2018ApJ...866..135V}, thus demanding an analytical framework to model the small-scale features of \hi.

One promising choice is the halo model of LSS \citep{2002PhR...372....1C}, which decomposes the description of a tracer into its distribution within a dark matter halo of given mass and the statistical properties of dark matter halos in our Universe. It has been shown to match observations of galaxy clustering for different color and luminosity (e.g. \citealt{2011ApJ...736...59Z}). For \hi, it allows constraints of astrophysics when compared with measurements of damped Lyman $\alpha$ systems (e.g. \citealt{2017MNRAS.464.4008P}) and \hi\ galaxies (e.g. \citealt{2019MNRAS.486.5124O}). Forecasts predict that \hi\ halo models of the \hi\ power spectrum will enable future constraints of astrophysics and cosmology (e.g. \citealt{2020MNRAS.496.4115C}).

In this paper, we examine in detail the prospects of using the \hi\ halo model to infer astrophysics from future intensity mapping observations. We verify the accuracy of the \hi\ halo model by using it to reconstruct the \hi\ autopower spectra from highly resolved, large-volume simulations of \hi\ in the low-redshift Universe \citep{2020MNRAS.493.5434S}. This is the first time such a comparison has been performed, building on the cross-correlation work in \cite{2019MNRAS.484.1007W} that considered only lognormal mock simulations. In this work, we consider $z\sim0.1$ and $z\sim1.0$, with the former being well studied through \hi\ galaxy surveys, and the latter being of more interest for cosmological observation. We forecast future observations by modelling survey sensitivities and explore the halo model parameter space with Monte-Carlo Markov Chain (MCMC). We consider the accuracy of using a parameterized empirical description of \hi\ properties, the constraining power of future surveys, the degeneracy of parameters, and the robustness of modelling. In particular, we investigate the power of extracting extra information from \hi\ shot noise in autocorrelations, similar to the idea of using the shot noise in cross-correlations described in \cite{2017MNRAS.470.3220W}.

This paper is organized as follows. The halo model formalism of calculating the \hi\ temperature power spectrum is discussed in Section \ref{sec:halomod}. In Section \ref{sec:sim}, we briefly describe the simulation we base our work on. In Section \ref{sec:hihalo}, we present the simulation and halo model calculation results of the \hi\ power spectrum. Calculation of \hi\ shot noise in the halo model is discussed in Section \ref{sec:SNform}. Statistical errors of the measurements are calculated in Section \ref{sec:det}. In Section \ref{sec:results}, we present the main result of this paper. In Section \ref{sec:SN}, we examine closely the viability of extracting extra information from shot noise. Finally, we make our concluding remarks in Section \ref{sec:conclusion}.

\section{\althi\ power spectrum calculation using {\tt halomod}}
\label{sec:halomod}
For the analytical calculation of the \hi\ brightness temperature power spectrum we use the open-source {\tt python} package {\tt halomod}\footnote{\url{https://github.com/steven-murray/halomod}} \citep{Murray2020}, which deals with halo model calculations. 
It is built upon (and closely related to) the {\tt hmf}\footnote{\url{https://github.com/steven-murray/hmf}} \citep{2013A&C.....3...23M} package, which calculates the required Halo Mass Function (HMF). 
{\tt halomod} provides a simple and intuitive interface for calculating the halo model power spectrum, and provides an extremely wide range of built-in models for many of the required subcomponents. 
Importantly, it also provides the flexibility to plug in custom models,
which we utilise for the \hi\ halo occupation distribution (HOD) and \hi\ density profile (see Section \ref{sec:sim}) to make a publicly available \hi\  extension\footnote{\url{https://halomod.readthedocs.io/en/latest/examples/extension.html}}.

{\tt halomod} splits the calculation of the tracer power spectrum into a set of fundamental modular components, each of which provides a number of implemented models:
\begin{itemize}
\item
An input cosmology to calculate mass variance $\sigma$ at various scales. Throughout this paper we set our fiducial cosmology to 1-year Wilkinson Microwave Anisotropy Probe (WMAP1) result \citep{2003ApJS..148..175S}, to be consistent with the simulation used in \cite{2020MNRAS.493.5434S} to measure the \hi\ power spectrum. The linear matter power spectrum $P_m(k)$ and growth rate are calculated using {\tt CAMB} \citep{2000ApJ...538..473L, 2002PhRvD..66j3511L}.
\item
A fitting formula to calculate the HMF, $n(m)$. Here we adopt the result of \cite{2008ApJ...688..709T} (hereafter  `Tinker08'), a modification to \cite{2001MNRAS.323....1S}.
\item
A fitting formula for scale-independent halo bias $b(m)$. Here we adopt the formulae of \cite{2010ApJ...724..878T} (hereafter `Tinker10'), a modification to \cite{1999MNRAS.308..119S}.
\item
An HOD for the tracer of interest, in our case the \hi\ HOD $\langle M_{\rm \hi}^{\rm cen,sat}(m)\rangle$ for both \hi\ in central galaxies (central \hi) and in satellite galaxies (satellite \hi) as described in Section \ref{sec:sim}.
\item
A tracer density profile $\rho_{\rm \hi}(r)$, as described in Section \ref{sec:sim}. The density profile enters the equations for calculating the power spectrum in the form of its normalised Fourier transform $u_{\rm \hi}(k)$.
\item
A concentration-mass (c-m) relation for tracer $c_{\rm \hi}(m)$, which relates the characteristic radius $r_{\rm s}$ to virial radius $r_{\rm vir}$ via $c_{\rm \hi}(m) = r_{\rm vir}/r_{\rm s}$. We choose the density profile of \cite{2007MNRAS.378...55M} with parameter values of \cite{2017MNRAS.464.4008P}.
\end{itemize}

The \hi\ brightness temperature autopower spectrum can be written as the sum of a two-halo, one-halo and shot noise term:
\begin{equation}
P_{\rm \hi}(k) = \overline{T}_{\rm \hi}^2\Big(P_{\rm 2h}(k)+P_{\rm 1h}(k)+P_{\rm SN}\Big).
\label{eq:p21}
\end{equation}

The two-halo term quantifies the correlation between pairs of \hi\ galaxies in different halos, while the one-halo term quantifies the correlation between pairs of \hi\ galaxies in the same halo.

The average brightness temperature $\overline{T}_{\rm \hi}$ can be calculated from the average \hi\ density:
\begin{equation}
    \bar{\rho}_{\rm \hi} = \int {\rm d}m\;n(m) \bigg[\langle M_{\rm \hi}^{\rm cen}(m) + M_{\rm \hi}^{\rm sat}(m)\rangle\bigg]
\end{equation}
via the relation $\overline{T}_{\rm \hi}=C_{\rm \hi}\bar{\rho}_{\rm \hi}$, with the conversion coefficient \citep{2006PhR...433..181F}:
\begin{equation}
    C_{\rm \hi}=\frac{3A_{12}h_{\rm P}c^3(1+z)^2}{32\pi m_{\rm H}k_{\rm B} \nu_{21}^2 H(z)}
\end{equation}
with $h_{\rm P}$ the Planck constant, $k_{\rm B}$ the Boltzmann constant, $m_{\rm H}$ the mass of the hydrogen atom, $A_{12}$ the emission coefficient of the 21-cm line transmission, and $\nu_{21}$ the rest frequency of the 21-cm emission. $H(z)$ is the Hubble parameter at redshift $z$.

The two-halo autopower spectrum $P_{\rm 2h}(k)$ is \citep{2019MNRAS.484.1007W}:
\begin{equation}
P_{\rm 2h}(k) =  b_{\rm \hi}^2(k) P_{\rm m}(k)
\end{equation}
with $b_{\rm \hi}$ being the \hi\ bias,
\begin{equation}
\begin{split}
    b_{\rm \hi}(k) =  \frac{1}{\bar{\rho}_{\rm \hi}}\int {\rm d}m\:n(m) b(m) \big[&\langle M_{\rm \hi}^{\rm cen}(m)\rangle\\
    &+\langle M_{\rm \hi}^{\rm sat}(m)\rangle u_{\rm \hi}(k|m)\big].
\end{split}
\label{eq:bhi}
\end{equation}
Note we define the \hi\ bias to be the square root of two-halo term of the \hi\ power spectrum over the matter power spectrum.

$P_{\rm 1h}(k)$ is the one-halo autopower spectrum:
\begin{equation}
\begin{split}
P_{\rm 1h}^{\rm dsc}(k) = \frac{2}{\bar{\rho}^2_{\rm \hi}}\int {\rm d}m\: n(m) \big[& u_{\rm \hi}(k|m) \langle M^{\rm sat}_{\rm \hi}(m)\rangle \langle M^{\rm cen}_{\rm \hi}(m)\rangle\\
&+ \frac{1}{2}\langle M^{\rm sat}_{\rm \hi}(m)\rangle^2 u_{\rm \hi}(k|m)^2
\big].
\end{split}
\label{eq:phi1h}
\end{equation}

The final component is the \hi\ shot noise $P_{\rm SN}$, which is assumed to be a scale-independent Poisson noise. We discuss this in detail in Section \ref{sec:SNform}.

For the given input, {\tt halomod} can efficiently calculate average brightness temperature, \hi\ power spectrum and shot noise.

\section{Simulation}
\label{sec:sim}
To construct our fiducial \hi\ power spectrum we use the results presented in \cite{2020MNRAS.493.5434S} which are based on the outputs from the GAlaxy Evolution and Assembly (GAEA) semi-analytic model (see \citealt{2014MNRAS.445..970D, 2016MNRAS.461.1760H, 2017MNRAS.469..968X} for detailed description of the model).
We select redshift $z\sim0.1$ and $z\sim1.0$ from the GAEA merger trees run on the 
 Millennium II (MII) large-scale dark matter cosmological simulation \citep{2009MNRAS.398.1150B} based on the WMAP1 cosmology \citep{2003ApJS..148..175S}. These redshifts are within the reach of both single dish and interferometric surveys, providing strong incentives for detailed investigation. The {\tt SUBFIND} algorithm \citep{Springel2001} is used to identify bound substructures (subhaloes) within standard friend-of-friend (FoF) dark matter haloes of the MII simulation. The most bound part of the FOF group hosts the central galaxy, while satellite galaxies are associated with all other bound subhaloes.
The GAEA model follows the evolution of these substructures and models complex interactions of gas reservoirs on them. Cold gas trapped in collapsing structures accretes into the central region and typically settles in a rotating disc, where star formation takes place. GAEA also models stellar feedback through, for example, supernova explosions, gas reheating and the generation of galactic outflows. The energy released by black holes at the centre of galaxies is considered as an additional source of reheating for the surrounding gas, preventing star formation [active galactic nuclei (AGN) feedback; see e.g. \citealt{Zoldan2019, 2019MNRAS.483.4922B, 10.1093/mnras/staa2251, 2020MNRAS.493.5434S}].
Crucial for our purposes, the GAEA model presented in \cite{2017MNRAS.469..968X} includes an on-the-fly partitioning of cold gas into its atomic (\hi) and molecular (H$_2$) components and an explicit dependence of star formation only on the latter. 
The partitioning is done by dividing the galactic disc in concentric rings and using an empirical power-law relation between the molecular to atomic ratio (H$_2$/\hi) and the hydrostatic mid-plane pressure of the disc \citep{2006ApJ...650..933B}.

The model is tuned to accurately match the local universe \hi\ mass function as measured in
\citet{2005ApJ...630....1Z} and \citet{2010ApJ...723.1359M}, using the blind \hi\ surveys \hi\ Parkes All Sky Survey (HIPASS; \citealt{Meyer2004}) and Arecibo Legacy Fast ALFA (ALFALFA; \citealt{Giovannelli2005}).

We note, however, that the model predicts a decrease of $\Omega_{\rm \hi}$ with redshift starting around $z \sim 1$ (see Fig. 4 of \citealt{2020MNRAS.493.5434S}), whereas absorption data suggest the opposite \citep{2020MNRAS.493.5854H}. This trend could be due to limitations of resolution or insufficient modelling (see discussion in \citealt{2020MNRAS.493.5434S}).

The simulation allows us to infer the total \hi\ mass in dark matter halos, known as an \hi\ HOD (see also \citealt{10.1093/mnras/stu445, 2017MNRAS.464.4008P,2018ApJ...866..135V}). 
Moreover, we can separate the contribution of the central galaxy from the satellite galaxies in each halo and reconstruct an \hi\ `density profile' for satellites.
The distinction between central and satellite galaxies in the \hi\ HOD has an impact on the \hi\ clustering properties \citep{2020MNRAS.493.5434S} and requires proper incorporation into the halo model, as discussed in Section \ref{sec:halomod}.

\begin{figure}
\centering
\includegraphics[width=0.49\textwidth]{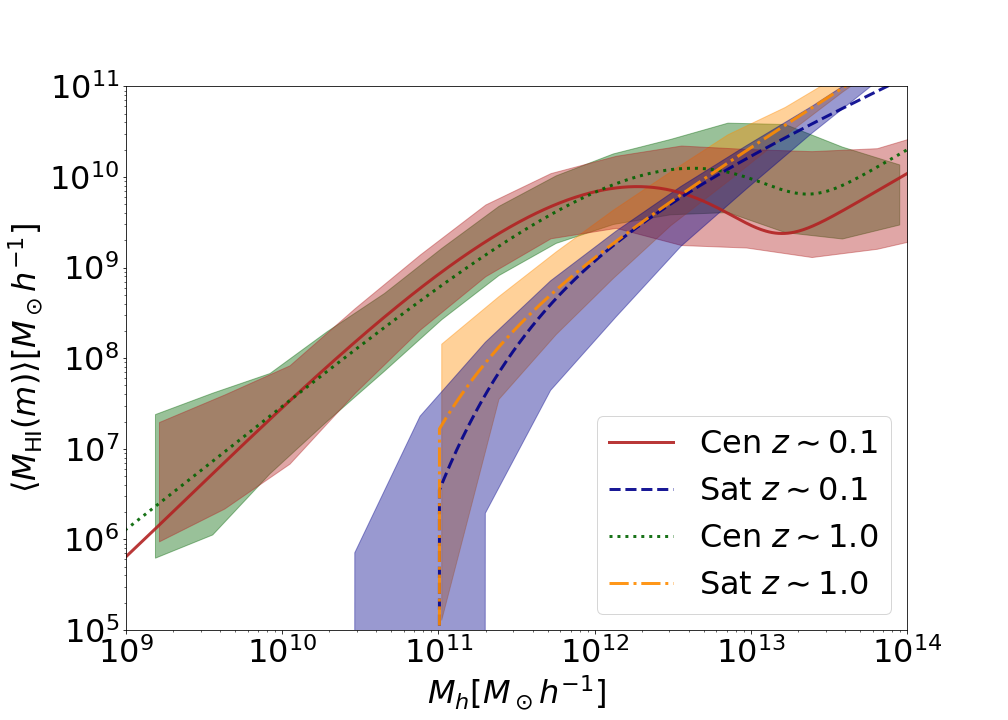}
\caption{The mean \hi\ mass-halo mass relation (\hi\ HOD) for central and satellite components from the GAEA simulation with its $1\sigma$ deviation (shaded areas), and our parametrization presented in Eqs. (\ref{eq:HODcen}) and (\ref{eq:HODsat}).}
\label{fig:HOD}
\end{figure}

In Fig. \ref{fig:HOD}, we report the \hi\ HOD from the GAEA simulation, with shaded regions depicting the scatter (1$\sigma$ confidence region) of \hi\ within halos, separating the \hi\ content of centrals and satellites. As discussed in \citet{2020MNRAS.493.5434S}, the total \hi\ mass in centrals
is suppressed in halos below a cut-off mass and follows a power-law relation up to halo mass $M_{\rm h}\sim 10^{12} M_\odot {h^{-1}}$. Above this halo mass, most likely the effect of AGN feedback causes a drop of the power-law behavior. For the most massive halos, we find $M_{\rm \hi}\propto M_{\rm h}$ . This behavior can be modelled following \citet{2020MNRAS.493.5434S} as
\begin{equation}
\begin{split}
\langle M_{\rm \hi}^{\rm cen}(M_{\rm h}) \rangle = M_{\rm h}& \Bigg[a_1^{\rm cen}\bigg(\frac{M_{\rm h}}{10^{10} M_\odot}\bigg)^{\beta_{\rm cen}} {\rm exp}\Big[{-\bigg(\frac{M_{\rm h}}{M^{\rm cen}_{\rm break}}\bigg)^{\alpha_{\rm cen}}}\Big] \\&+a_2^{\rm cen}\Bigg] {\rm exp}\left[{-\bigg(\frac{M_{\rm min}^{\rm cen}}{M_{\rm h}}\bigg)^{0.5}}\right]
\end{split} 
\label{eq:HODcen}
\end{equation}
extending the parametrization of \citet{2019MNRAS.483.4922B}.

For the total \hi\ content in satellites as a function of halo mass, the GAEA simulation shows a simpler relation: a power law with a low-mass cut-off, due to the lack of satellite galaxies in low-mass halos ($M_{\rm h} < 10^{11} M_\odot { h^{-1}}$). We follow again \citet{2020MNRAS.493.5434S} and use
\begin{equation}
\langle M_{\rm \hi}^{\rm sat}(M_{\rm h}) \rangle = 
M_0^{\rm sat}\bigg( \frac{M_{\rm h}}{M^{\rm sat}_{\rm min}}\bigg)^{\beta_{\rm sat}}
{\rm exp}\left[{-\bigg(\frac{M^{\rm sat}_{\rm min}}{M_{\rm h}}\bigg)^{\alpha_{\rm sat}}}\right].
\label{eq:HODsat}
\end{equation}
\begin{table*}
\centering

\begin{tabular}{ccccccccccc}
Parameters &  $a_1^{\rm cen}[10^{-3}]$   & $a_2^{\rm cen}[10^{-4}]$ & $\alpha_{\rm cen}$ & $\beta_{\rm cen}$ & ${\rm log_{10}}\big[M^{\rm cen}_{\rm break}\big]$ & ${\rm log_{10}}\big[M_{\rm min}^{\rm cen}\big]$  & ${\rm log_{10}}\big[M_0^{\rm sat}\big]$ & $\alpha_{\rm sat}$ & $\beta_{\rm sat}$ & ${\rm log_{10}}\big[M^{\rm sat}_{\rm min}\big]$ \\ \hline
$z\sim0.1$ & $4.66$ & $1.09$ & $0.41$ & $0.85$ & $10.66$ &  -1.99 & $9.51$ & 0.70 & $0.81$ & $12.00$\\
$z\sim1.0$ & $3.00$ & $2.00$ & $0.56$ & $0.43$ & $11.86$ &  -2.99 &  $8.58$ & 0.84 & $1.10$ & $11.40$ 
\\ \hline
\end{tabular}
\caption{Fiducial values for \hi\ HOD parameters from our simulation for $z\sim 0.1$ and $z\sim 1.0$.}
\label{tab:parsfid}
\end{table*}

The best-fitting values of the parameters of Eqs. \ref{eq:HODcen} and \ref{eq:HODsat} are
presented in Table \ref{tab:parsfid} and overplotted on the simulation result in Fig. \ref{fig:HOD}. These will be used as our fiducial values to be reconstructed in the framework of the halo model.

We also need to model how the \hi\ content in satellites is distributed within the halo, i.e. their density profile. We are interested in its overall shape $\Tilde{\rho}_{\rm \hi}(r)$, normalized at some length scale $r_0$.

\begin{figure}
\centering
\includegraphics[width=0.49\textwidth]{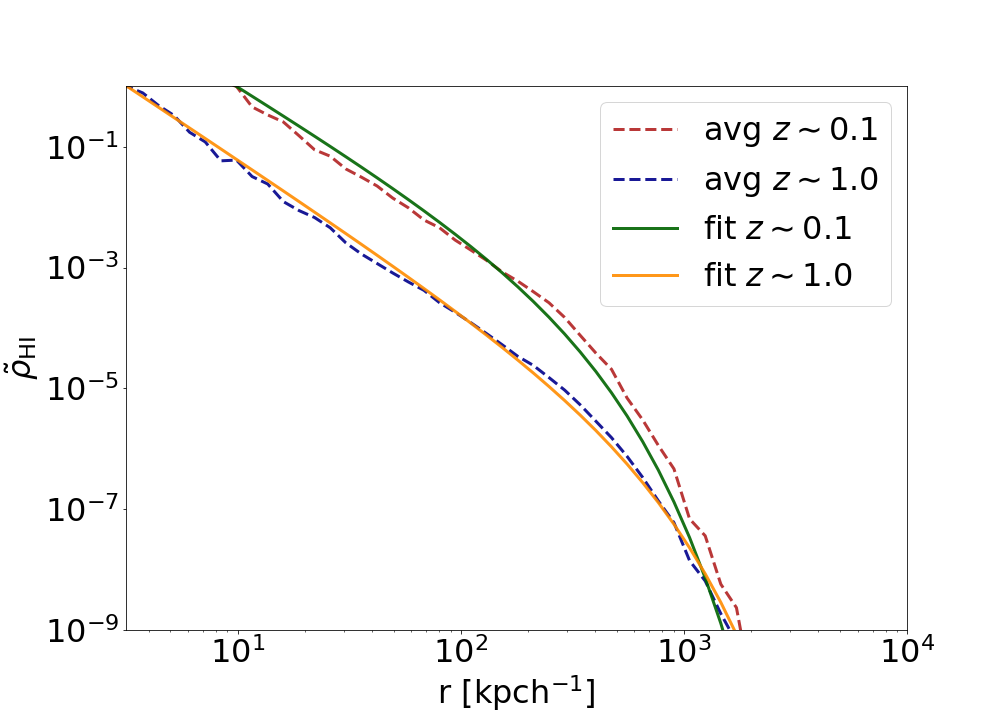}
\caption{The {renormalized} \hi\ density profile. Dashed line `avg' is the results calculated from the GAEA simulation, averaged around $m\sim 10^{12} M_\odot h^{-1}$, and `fit' is our fitting formulae using Eq. (\ref{eq:HIdp}). For different redshift the density profile is renormalized at different scales for better illustration. }
\label{fig:HIdp}
\end{figure}

As shown in Fig. 9 of \cite{2020MNRAS.493.5434S}, the average density profile varies as a function of halo mass. However, the clustering of \hi\ is primarily determined by halos of $M_{\rm h} \sim 10^{12} M_\odot {h^{-1}}$, hence for the density profile we concentrate on the average \hi\ density profile for halos around this mass. Additionally, the shape of the density profile does not change much with halo mass, and can be well described as an average. In Fig. \ref{fig:HIdp}, we show the density profile averaged over halos from mass $10^{11} M_\odot { h^{-1}}$ to $10^{13} M_\odot { h^{-1}}$, renormalized at $10\,{\rm kpc\,} h^{-1}$ for $z\sim0.1$ and $1\,{\rm kpc\,} h^{-1}$ for $z\sim1.0$. 

The density profile follows a power law at small scales, and then drops exponentially beyond a certain radius $r$. Our result is consistent with \cite{2017MNRAS.464.4008P} and  \cite{2018ApJ...866..135V}, that suggest an extremely concentrated profile, either cored Navarro-Frenk-White (cored-NFW; \citealt{2004MNRAS.355..694M}) or a power law. We parametrize the profile as
\begin{equation}
\rho_{\rm \hi} = \rho_{\rm s} \bigg(\frac{r_s}{r}\bigg)^b {\rm exp}\bigg[-a \frac{r}{r_{\rm s}}\bigg]
\label{eq:HIdp}
\end{equation}
where $a$ and $b$ are free parameters, and $r_{\rm s}\equiv r_{\rm vir}/c(m)$ is a characteristic scale radius (with $c(m)$ a so-called `concentration' parameter and $r_{\rm vir}$ the virial radius). 

The power spectrum calculation is dependent on the {Fourier transform} of the profile, whose analytical expression reads
\begin{equation}
\begin{split}
u(k|m)=&\frac{-1}{1+(K/a)^2}\Bigg(\bigg(1+K^2/a^2\bigg)^{b/2}\\
&\times\Gamma(2-b)\sin\bigg[(b-2){\rm arctan}\big[K/a\big]\bigg]\Bigg),(b\ne 2)
\end{split}
\label{eq:HIdpFT}
\end{equation}
where $K \equiv k r_{\rm s}$ and $\Gamma$ is the Gamma function.

Our fitting parameters are $a = 0.040,b = 2.262$ for $z\sim0.1$ and $a = 0.049,b = 2.248$ for $z\sim1.0$ upon choosing the input c-m relation from \cite{2007MNRAS.378...55M} with fiducial values from \cite{2017MNRAS.469.2323P}. The fitted results are plotted in Fig. \ref{fig:HIdp}.

\section{Reconstructing the \althi\ Power Spectrum}
\label{sec:hihalo}

In this section, we discuss the simulation results of \cite{2020MNRAS.493.5434S} in the context of {\tt halomod} and describe our choices for the halo model inputs presented in Section \ref{sec:halomod}. 
\subsection{Halo Mass Function}
The halo mass function (HMF) $n(m)$ describes the number density of dark matter haloes of any given mass. It has been extensively studied in the context of simulating LSS (e.g. \citealt{1974ApJ...187..425P, 2001MNRAS.323....1S, 10.1046/j.1365-2966.2003.07113.x, 2008ApJ...688..709T, 2011ApJ...732..122B}), and is believed to be relatively universal and thus insensitive to specific values of cosmological parameters in $\Lambda$CDM.

We use the `Tinker08' HMF that matches the MII simulation results well \citep{2008ApJ...688..709T}.

As the resolution limit of the simulation is around $M_{\rm h} \sim 10^{10}M_{\odot} h^{-1}$, we introduce a halo mass cut-off such that the halo number density derived from the HMF matches the one from the simulation. We measure $n_{\rm h} \approx 0.31 [{\rm Mpc^{-3}}h^3]$ for $z\sim0.1$ and $n_h \approx 0.33 [{\rm Mpc^{-3}}h^3]$ for $z\sim1.0$ that translate into halo mass cut-off at $m_{\rm min}= 10^{9.72}M_{\odot}{ h^{-1}}$ for $z\sim0.1$ and $m_{\rm min}= 10^{9.81}M_{\odot}{h^{-1}}$ for $z\sim1.0$.

\subsection{Halo Bias}
The halo bias is defined as the ratio of the two-point correlation function of dark matter halos of a given mass, and the dark matter two-point correlation function. As a first-order approximation it can be considered as a linear scale-independent bias, $b(m)$, determined solely by the mass of the halo.

We use the halo bias formula from `Tinker10' that fits the MII simulation well \citep{2010ApJ...724..878T}. The halo autopower spectrum can be computed as
\begin{equation}
P_{\rm hh}^{\rm lin}(k) = \int {\rm d}m_1\;{\rm d}m_2\;b(m_1)b(m_2)n(m_1)n(m_2)P_{\rm lin}(k).
\label{eq:phh}
\end{equation}
We emphasize that this is only true for large scales where the linear approximation still holds. 

\begin{figure}
\centering
\includegraphics[width=0.49\textwidth]{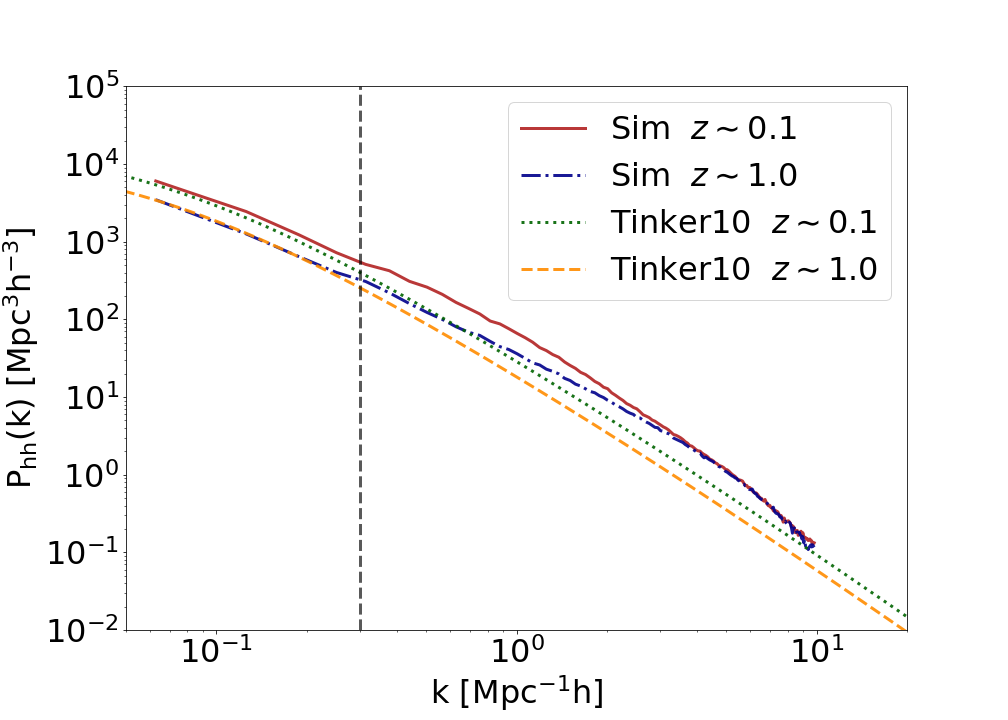}
\caption{The halo autopower spectrum from our simulation (`Sim') and from {\tt halomod} (`Tinker10') according to Eq. (\ref{eq:phh}) at $z\sim0.1$ and $z\sim1.0$. The vertical dashed line shows the scale where linearity breaks.}
\label{fig:halobias}
\end{figure}

In Fig. \ref{fig:halobias}, we present the halo autopower spectrum calculated using the `Tinker10' model of halo bias, comparing against the power spectrum from our simulation. Note that linearity breaks for 
$k>k_{\rm NL}$ with $k_{\rm NL}\sim 0.3 {\rm Mpc^{-1}}h$.

\subsection{\althi\ HOD and Density Profile}
In Section \ref{sec:sim}, we discussed the \hi\ HOD fitting formulae and the \hi\ density profile obtained in \citet{2020MNRAS.493.5434S} using the GAEA simulation.
However, in describing the \hi\ power spectrum within the context of the halo model, some of the parameters of Eqs. (\ref{eq:HODcen}) and (\ref{eq:HODsat}) are either degenerate or ineffective on the power spectrum. Therefore, we simplify the \hi\ HOD formulae fixing the following parameters: $[M_{\rm min}^{\rm cen}, a_2^{\rm cen}, \alpha_{\rm  cen}, \alpha_{\rm  sat}]$. The cut-off mass $M_{\rm min}^{\rm cen}$ is physically motivated (e.g. \citealt{2018ApJ...866..135V}) but difficult to see its effects in the GAEA simulations due to the low resolution. Since the results of the fit suggest that the cut-off in equation (\ref{eq:HODcen}) is effectively unity, we do not consider it in our final model.

The parameter $a_2^{\rm cen}$ is usually too small to have an impact on the power spectrum and controls the \hi\ mass of extremely massive halos, which are not abundant enough to have a sizeable effect on the power spectrum.
Finally, $\alpha_{\rm cen}$ and $M_{\rm break}^{\rm cen}$, and likewise  
$\alpha_{\rm sat}$ and $M^{\rm cen}_{\rm break}$ are degenerate thus we keep only one of each.

Our resulting model parameter set, i.e. $[a_1^{\rm cen},\beta_{\rm cen}, M^{\rm cen}_{\rm break}, M_0^{\rm sat}, \beta_{\rm sat}, M^{\rm sat}_{\rm min}]$, has a significant impact on the resulting power spectrum and regulates the overall \hi\ density $\Omega_{\rm HI}$ and a scale-dependent \hi\ bias $b_{\rm \hi}(k)$. 

While the \hi\ energy density value at $z\sim0.1$ of the semi-analytical model is in agreement with local Universe emission observations, when we compute the value of $\Omega_{\rm \hi}$ using the halo model we find a slightly lower value. This is due to the small differences in amplitude between the `Tinker08' HMF and the actual HMF from simulation, as well as other secondary effects such as correlation between the scatter of HMF and scatter of \hi\ HOD.

The mismatch is a natural outcome of the limited resolution and volume of the simulation, as well as the limitations of ignoring assembly bias (dependency of halo properties other than halo mass; \citealt{2007MNRAS.377L...5G}). A detailed analysis of the halo model in terms of these shortcomings is outside the scope of this work. We simply rescale the overall \hi\ density to recover agreement with observations.
We rescale also the $\Omega_{\rm \hi}$  value at $z\sim 1$. Instead of the GAEA value, we use the measurement $\Omega_{\rm \hi}b_{\rm \hi} = 0.63\times 10^{-3}$ from \citet{10.1093/mnrasl/slt074} and  \cite{2013ApJ...763L..20M}. This is in agreement with absorption results at this redshift \citep{2017MNRAS.471.3428R} and allows to overcome the possible limitations discussed in Section \ref{sec:sim}.
We stress, however, that the GAEA model used here produces results in good agreement with most common parametrizations and that its fairly realistic modelling for the \hi\ is enough for the purposes of testing the \hi\ halo model and its constraining power. \\
To summarize, we use
\begin{align}
\Omega_{\rm HI} = 0.43\times 10^{-3},z\sim0.1\\
\Omega_{\rm HI} = 0.46\times 10^{-3},z\sim1.0.
\label{eq:omegabhi}
\end{align}


Similarly the aforementioned setbacks also affect the calculation of $b_{\rm \hi}$, and therefore, the amplitude of the \hi\ power spectrum from the analytical expressions on the halo model still does not accurately match the GAEA result with enough precision for our study. Since it is beyond the scope of this paper to fully examine and compare these two, we simply match the overall amplitude at large scales, rescaling the value of the first $k$ bin at $0.07 {[\rm Mpc^{-1}}h]$. Note that this has no effect on non-linear scales.

In the following, we discuss the impact of the parameters on the \hi\ power spectrum. $a_1^{\rm cen}$ and $M_0^{\rm sat}$ control the central and satellite fraction, respectively. If the central fraction $a_1^{\rm cen}$ increases, the satellite fraction will drop in order to keep the overall \hi\ density constant. Reducing the satellite fraction $M_0^{\rm sat}$ lowers the \hi\ power spectrum on small scales since the small-scale power depends on how the satellite \hi\ distributes within halos. 

$\beta_{\rm cen}$ and $\beta_{\rm sat}$ control the fraction of \hi\ in halos of different mass. Increasing $\beta_{\rm cen,sat}$, more \hi\ is painted in 
high-mass halos resulting in a higher \hi\ power spectrum at large scales and less power at intermediate and small scales.

The effect of $M^{\rm cen}_{\rm break}$ on the \hi\ HOD model is related to the AGN feedback and this naturally defines the lower and upper limit for this parameter. If $M^{\rm cen}_{\rm break}$ is too small, the \hi\ HOD model effectively becomes $a_2^{\rm cen} M_{\rm h}$, resulting in an improper description of the  \hi\ content of intermediate- and low-mass halos.
On the other hand, if $M^{\rm cen}_{\rm break}$ is too large, it will impact only the very high mass halos that are too scarce to have a sizeable contribution in the \hi\ power spectrum. 
Similar arguments also apply to $M^{\rm sat}_{\rm min}$. 

The \hi\ density profile models the distribution of satellite \hi\ within halos. In our case, it is described by two parameters (see equation \ref{eq:HIdp}), the exponential cut-off $a$ and power-law index $b$. Since $a$ only affects the satellite density profile on very large $r$ where two-halo term is likely to dominate, we keep $a$ fixed and keep $b$ as a free parameter.

 The parameter $b$ affects how satellite \hi\ is distributed within halos. Consequently, as $b$ increases \hi\ is more concentrated, and thus amplifies small-scale \hi\ clustering.

\subsection{Non-Linear Effects}
The standard halo model assumes that the halo centres trace the {linear} matter overdensity, and therefore predicts a halo autopower spectrum proportional to the {linear} dark matter power spectrum. As seen in Fig. \ref{fig:halobias}, this assumption only holds for $k<0.3\,{\rm Mpc^{-1}}h$. Assuming that the one-halo term accurately describes the features of the \hi\ distribution on small scales, failing to account for non-linear effects will significantly bias the measurement of model parameters.

In order to accurately compute the non-linear matter power spectrum before applying the halo model, methods such as the Effective Field Theory of Large Scale Structure (EFTofLSS; \citealt{2012JCAP...07..051B, 2012JHEP...09..082C}) and the Effective Halo Model \citep{2020arXiv200409515P} can be used. Comparing and using different methods to accurately calculate the underlying halo clustering is beyond the scope of this paper. Here, since we are interested in extracting astrophysics and keep cosmology fixed, we simply adopt an effective `transfer function':
\begin{equation}
\tilde{T}(k)\equiv P_{\rm hh}(k)/P_{\rm hh}^{\rm linear}(k).
\label{eq:tk}
\end{equation}
The temperature power spectrum of Eq. (\ref{eq:p21}) then becomes
\begin{equation}
P_{\rm \hi}(k) = \bar{T}_{\rm HI}^2\Big(\tilde{T}(k)P_{\rm 2h}(k)+P_{\rm 1h}(k)+P_{\rm SN}\Big)
\label{eq:nonlinearp21}
\end{equation}
which includes an effective correction to the two-halo term to account for non-linearity.

\begin{figure}
\centering
\includegraphics[width=0.49\textwidth]{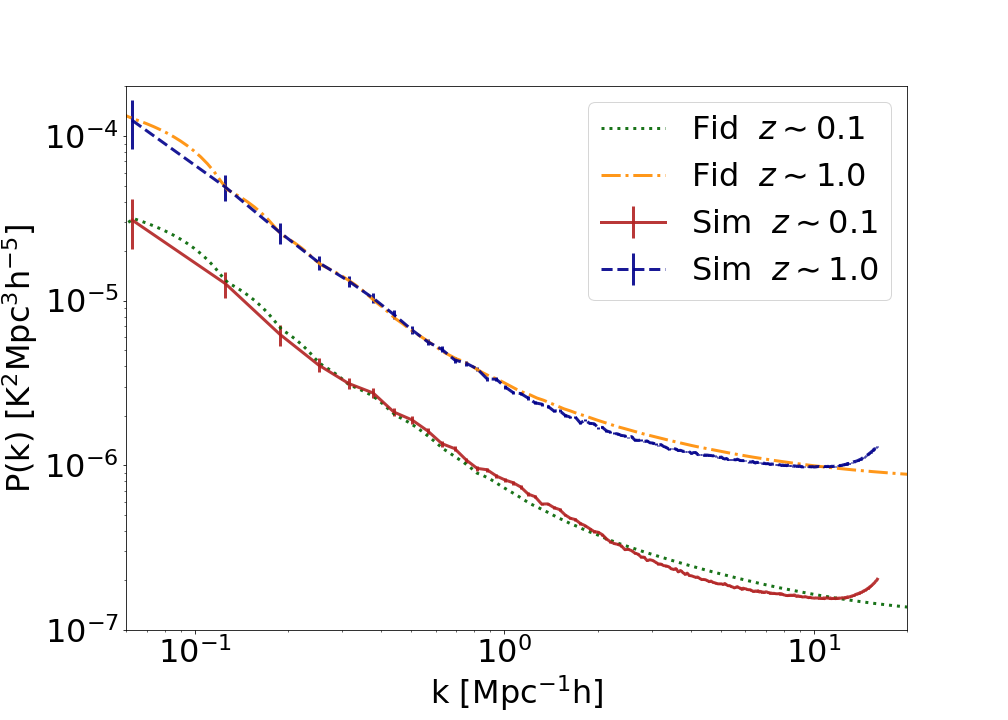}
\caption{The temperature power spectrum, from simulation (`Sim'), and power spectrum with non-linear correction according to Eqs. (\ref{eq:tk}) and (\ref{eq:nonlinearp21}), which we choose as the fiducial power spectrum for fitting (`Fid'). A shot noise estimated using Eq. (\ref{eq:PSNsim}) has been added. Error bars are estimated as the cosmic variance from the simulation box.}
\label{fig:nonlinearps}
\end{figure}

In Fig. \ref{fig:nonlinearps}, we present the non-linear corrected power spectrum of the \hi\ brightness temperature. The non-linear halo clustering spectrum is set to be equal to the results from simulation. This correction considerably improves the accuracy of reproducing the temperature power spectrum at intermediate scales. As the correction is only necessary for the positions of halos, the modelling of \hi\ clustering from intermediate to small scales still solely relies on the halo model.

We use the aforementioned formalism to obtain a power spectrum from  {\tt halomod} that is in agreement with the power spectrum measured on the GAEA simulation up to $k \sim 10\;{\rm Mpc^{-1}}h$. Nevertheless, as we will discuss in Section \ref{sec:det}, future observational constraints are forecasted to go to much smaller scales, $k\gtrsim100 {\rm Mpc^{-1}}h$. To test the ability of {\tt halomod} to extract the astrophysical information contained also in these small scales, we then retain as a fiducial \hi\ power spectrum our model presented in Eq. (\ref{eq:nonlinearp21}),  calibrated on the GAEA simulation but extending up to much smaller scales.

\section{\althi\ Shot Noise}
\label{sec:SNform}
\subsection{\althi\ Shot Noise in Halo Model}
In this section, we discuss in detail how to estimate shot noise in the context of the halo model, and how it can be utilized for parameter inference (see also \citealt{2019MNRAS.484.1007W} for a discussion).

In galaxy surveys, the shot noise of the clustering is $P_{\rm SN}\sim 1/n_g$, where $n_g$ is the number density of galaxies. For the \hi\ power spectrum, each source is weighted by \hi\ mass such that
\begin{equation}
    P_{\rm SN} = V\frac{\Big\langle\sum\limits_i \big(M_{\rm \hi}^{i}\big)^2\Big\rangle}{\Big(\big\langle \sum\limits_i M_{\rm \hi}^{i}\big\rangle\Big)^2}
\label{eq:PSNsim}
\end{equation}
where $i$ runs over all \hi\ sources within volume $V$. In the context of intensity mapping, information on \hi\ sources is only accessible through simulations.

It has been suggested that the $k\rightarrow0$ limit of the one-halo term $P_{\rm SN}=P_{1{\rm h}}(k=0)$ may be used as an estimation of the shot noise (e.g. \citealt{2018ApJ...866..135V}). This implies that halos are the discrete sources of \hi. This is inaccurate for our simulation since the discrete sources of \hi\ are not halos but galaxies within. Using this formalism will underestimate the number density of \hi\ sources, and thus overestimate shot noise. In our case, we report that the one-halo term gives a shot noise that is about four (two) times larger than the fiducial shot noise simulation at $z\sim0.1$ ($z\sim1$). 

To accurately estimate the \hi\ shot noise in the context of the halo model, one needs to know the galaxy HOD to know the number density of \hi\ sources \citep{2019MNRAS.484.1007W}:
\begin{equation}
P_{\rm SN}^{\rm no\;scatter} = \int {\rm d}m\; n(m) \sum_k^{\rm cen,sat} \langle M_{\rm \hi, field}^k (m)\rangle^2 \langle N_{\rm field}(m)\rangle /\bar{\rho}_{\rm \hi}^2
\label{eq:snfield}
\end{equation}
where $\langle M_{\rm \hi, field}^k (m)\rangle$ is the ensemble average of \hi\ mass {\it per galaxy} and $\langle N_{\rm field}(m)\rangle$ is the galaxy HOD, the mean number of galaxy as a function of the hosting halo mass. Here it is assumed that the \hi\ in a particular halo is equally distributed within each galaxy, so that $\langle M_{\rm \hi, field}^k (m)\rangle = \langle M_{\rm \hi}^k/N_{\rm g}^k\rangle$. 

From Fig. \ref{fig:HOD}, one can see that the variance of the \hi\ HOD relation
produced from our simulation is quite large. A similar scatter is also predicted in other simulations (e.g. \citealt{2018ApJ...866..135V}). This variation leads to a non-negligible effect on the amplitude of the \hi\ shot noise that we derive in detail in Appendix \ref{app:sn}. The result yields
\begin{equation}
\begin{split}
    P_{\rm SN} = \frac{1}{\Bar{\rho}_{\rm \hi}^2}\sum_{k}^{\rm cen,sat}
    \int {\rm d}m\; &n(m) \big\langle M_{\rm \hi}^k(m)\big\rangle^2 \big\langle N_g^k(m)\big\rangle^{-1}\\&\times\Big(1+\big(\sigma_{\rm \hi}^k(m)\big)^2 \Big)
\label{eq:PSN}
\end{split}
\end{equation}
where $\sigma_{\rm \hi}^k(m)$ is the scatter of \hi\ mass per galaxy in halos, i.e., the renormalized standard deviation of \hi\ mass per galaxy in halos of given mass (see Eq. \ref{eq:std}). For simplicity, from here on we refer to $\sigma_{\rm \hi}^k(m)$ as \hi-HOD scatter, as these two are tightly linked. For central galaxies, this scatter directly corresponds to the scatter of $\langle M_{\rm \hi}^{\rm cen}(m)\rangle$. For satellite galaxies, it can be related to the scatter of $\langle M_{\rm \hi}^{\rm sat}(m)\rangle$ as shown in Eq. (\ref{eq:hodfield}).

We report an estimation of 43 $[{\rm Mpc^{-3}}h^3]$ at $z\sim0.1$ and 52 $[{\rm Mpc^{-3}}h^3]$ at $z\sim1.0$ for \hi\ shot noise using fiducial parameter values, consistent with shot noise measured from our simulations, which is 47$[{\rm Mpc^{-3}}h^3]$ at $z\sim0.1$ and 61$[{\rm Mpc^{-3}}h^3]$ at $z\sim1.0$. At both redshifts the difference between estimation and measurement is around 10\%. With this formalism, we can potentially measure the \hi-HOD scatter in galaxies from future \hi\ intensity mapping observations. 

\subsection{\althi\ Shot Noise in terms of \althi\ Mass Function}
Conventional \hi\ galaxy surveys measure the distribution of \hi\ galaxies in terms of the \hi\ Mass Function (\hi\ignorespaces MF; \citealt{2005MNRAS.359L..30Z,2010ApJ...723.1359M}). The \hi\ignorespaces MF $\phi_{\rm \hi}$ quantifies the number density of \hi\ galaxies as a function of their \hi\ mass $n_{\rm \hi}$ in logarithmic mass bins \citep{1990AJ....100..999B}. We can re-formulate the \hi\ shot noise in  Eq. (\ref{eq:PSNsim}) and divide galaxies into \hi\ mass bins using the same formalism: 

\begin{equation}
P_{\rm SN} = \frac{\int {\rm d\,log}M_{\rm \hi}\; \phi_{\rm \hi}\big(M_{\rm \hi}\big) M_{\rm \hi}^2}{\bigg(\int {\rm d\,log}M_{\rm \hi}\; \phi_{\rm \hi}\big(M_{\rm \hi}\big)M_{\rm \hi}\bigg)^2}
\label{eq:himf}
\end{equation}

We report that using the \hi\ignorespaces MF measured from the \hi\ galaxy catalogue from simulation (see Fig. 2 of \citealt{2020MNRAS.493.5434S}), the above equation gives a shot noise of 47${\rm Mpc^{-3}}h^3$ for $z\sim0.1$ and 60${\rm Mpc^{-3}}h^3$ for $z\sim1.0$. These estimates match the fiducial values with 1\% level deviation. 

The \hi\ignorespaces MF is usually parametrized with a Schechter function \citep{1976ApJ...203..297S}:
\begin{equation}
\phi_{\rm \hi}\big(M_{\rm \hi}\big)\equiv \frac{{\rm d}n_{\rm \hi}}{{\rm dlog} M_{\rm \hi}} = {\rm ln}10 \: \phi_* \Big(\frac{M_{\rm \hi}}{M_*}\Big)^{\alpha+1}e^{-\frac{M_{\rm \hi}}{M_*}}
\end{equation}
A Schechter function has three free parameters, which means to fully determine the Schechter function, one needs to measure at least three observables. The \hi\ignorespaces MF describes the density of galaxies with respect to their \hi\ mass, and we can relate to the number density of \hi\ galaxies $n_{\rm g}$ as
\begin{equation}
n_{\rm g} = \int {\rm d}M_{\rm \hi}\; n_{\rm \hi}\big(M_{\rm \hi}\big)
\label{eq:ng}
\end{equation}
and the density of \hi\ $\Omega_{\rm HI}$ as
\begin{equation}
\Omega_{\rm \hi} = \int {\rm d}M_{\rm \hi}\; n_{\rm \hi}\big(M_{\rm \hi}\big)M_{\rm \hi}\big/\rho_c
\label{eq:omegahimf}
\end{equation}
where $\rho_c$ is the critical density of the Universe.

In Section \ref{sec:SN}, we combine these three observables $n_{\rm g}$, $\Omega_{\rm \hi}$, and $P_{\rm SN}$ of  Eq.~(\ref{eq:himf}) to constrain the parameters of the \hi\ignorespaces MF.

\section{Detectability}
\label{sec:det}
In this section, we present the detectability of the \hi\ power spectrum by future interferometric intensity mapping surveys following descriptions in \cite{2006ApJ...653..815M,2015ApJ...803...21B} and \cite{2017MNRAS.470.3220W}. 

Unlike single-dish surveys  where the dish diameter determines the observable perpendicular scales $k_\perp$, for interferometric observations, the relative distances between the dishes of the array and the resulting baseline density $n(u)$ in visibility space determine the $k$ range of the power spectrum from the relation $k_\perp\equiv 2\pi u /\chi$, where $\chi$ is the comoving radial distance at the observed redshift.

The resulting receiver noise is a function of scale given by
\begin{equation}
\sigma_{\rm T} (k_\perp) = \frac{\lambda^2 T_{\rm sys}}{A_{\rm dish}\sqrt{\Delta \nu n(u) {\rm d}^2u t_{\rm int} N_{\rm beam} N_{\rm pol}}}
\label{eq:sigmaT}
\end{equation}
where $A_{\rm dish}\approx \pi D_{\rm dish}^2/4$ is the effective collecting area of one dish, $t_{\rm int}=t_{\rm tot}\Omega_{\rm FOV}/\Omega_{\rm surv}$ is the integration time with total observation time of $t_{\rm tot}$ and total survey area of $\Omega_{\rm surv}$, $N_{\rm beam}$ the number of beams. and $N_{\rm pol}$ the number of polarizations per dish. The system temperature  $T_{\rm sys}= T_{\rm sky}+T_{\rm inst}$ is the combined temperature of sky and instrument. We model $T_{\rm sky} = 60{\rm K} \big(300 {\rm MHz}/\nu\big)^{2.55}$. 

The baseline density $n(u)$ is renormalized to the total number of baselines:
\begin{equation}
\int n(u) {\rm d}^2 u = N_{\rm dish}(N_{\rm dish}-1)/2
\label{eq:renormnu}
\end{equation}
The noise power spectrum can then be written as
\begin{equation}
P_{\rm N}(k_\perp) = \sigma_{\rm T}^2(k_\perp)V_{\rm pix} {\rm d}^2u/\Omega_{\rm FOV}
\end{equation}
where ${\rm d}^2u=2\pi u {\rm d}u$ is the 2D pixel in visibility space, $\Omega_{\rm FOV}\equiv \lambda^2/A_{\rm dish}$ is the field of view of the observation with an effective collecting area $A_{\rm dish}$. $V_{\rm pix}$ is the 3D voxel volume corresponding to the field of view:
\begin{equation}
V_{\rm pix}=r^2\Omega_{\rm FOV} r_\nu \frac{\Delta\nu}{\nu}
\end{equation}
where $r_\nu\equiv \frac{c(1+z)}{H(z)}$, $\nu$ is the observed frequency, and $\Delta\nu$ is the channel frequency width.

We average the noise power spectrum in $k$-spheres by computing
\begin{equation}
    P_{\rm N}(k) = \int \frac{k_\perp P_{\rm N}(k_\perp)}{k\sqrt{k^2-k^2_\perp}}{\rm d}k_\perp
\end{equation}
The power spectrum is measured multiple times in the Fourier plane for a $k$-bin of $\Delta k$, and thus the noise of the power spectrum measurement is:
\begin{equation}
\sigma_{\rm P}(k) = \frac{1}{\sqrt{N_{\rm mode}}} (P_{\rm \hi}(k) + P_{\rm N}(k))
\end{equation}
Note that cosmic variance has been included, assuming a Gaussian sample variance of $P_{\rm \hi}(k)/\sqrt{N_{\rm mode}}$, with the number of modes $N_{\rm mode}$ being
\begin{equation}
N_{\rm mode} = \frac{V_{\rm surv}}{(2\pi)^3}2\pi k^2 \Delta k
\end{equation}
in terms of the survey volume $V_{\rm surv}$.
\begin{table*}
\centering
\begin{tabular}{ccccccccc}
Survey & Redshift & $N_{\rm dish}$ & $N_{\rm beam}$  & $N_{\rm pol}$ & $T_{\rm inst}$[K] & $D_{\rm dish}$[m] & $t_{\rm tot}$[h] & $\Omega_{\rm surv}$[deg$^2$]    \\ \hline
DINGO & [0.07, 0.11] & 36 & 30 & 2 & 90 & 12.0 & 500 & 150\\
MIGHTEE & [0.07, 0.11] & 64 & 1 & 2 & 29 & 13.5 & 1000 & 20\\
SKA-MDB2 & [0.07, 0.11] & 190 & 1 & 2 & 28 & 15.0 & 10000 & 5000\\
SKA-WB1 & [0.9, 1.1] & 190 & 1 & 2 & 28 & 15.0 & 10000 & 20000\\
SKA-DB1 & [0.9, 1.1] & 190 & 1 & 2 & 28 & 15.0 & 10000 & 100\\
\end{tabular}
\caption{Survey and array specifications considered in our analysis.}
\label{tab:survey}
\end{table*}

For survey strategies, we consider two ongoing SKA pathfinder surveys, the `deep' field of the Deep Investigation of Neutral Gas Origins (DINGO) survey \citep{2012MNRAS.426.3385D} at the ASKAP at $0<z<0.26$ , and the MeerKAT International GHz Tiered Extragalactic Exploration (MIGHTEE) survey \citep{2016mks..confE...6J} using MeerKAT at $0<z<0.36$, as well as various planned SKA1-MID surveys. In this study, we use a medium redshift of $z\sim0.1$ for the comparison. Both pathfinder surveys are deep observations with relatively small areas as listed in Table \ref{tab:survey}. 

For SKA-MID, we forecast the Medium-Deep Band 2 Survey (from now on referred to as SKA-MDB2) as discussed in \cite{2020PASA...37....7S}, which covers 5000 deg$^2$ in 10000 hours. For $z\sim1.0$, we use the Wide Band 1 Survey (from now on referred to as SKA-WB1) that covers 20000 deg$^2$ in 10000 hours. However, since it is relatively shallow, we also suggest a Deep Band 1 Survey (from now on referred to as SKA-DB1) with a total observation time of 10000 h and an area of 100 deg$^2$. 

The observed redshift range for these arrays is larger than the one we examine in our forecasts. However, treating the entire survey volume as one effective redshift when the range is large leads to non-trivial effects coming from the redshift evolution of \hi\ along the line of sight, known as the light-cone effect \citep{2012MNRAS.424.1877D, 2014MNRAS.442.1491D}. Thus, we limit our study to a smaller redshift range where light-cone effects are more likely to be trivial. Moreover, the redshift range used here is enough to provide precise measurement, and in future experiments, using multiple redshift bins will determine the evolution of \hi.

For the MIGHTEE forecast, we use the baseline density derived from a MIGHTEE observational 2d visibility coverage of an 11.2 h tracking of the Cosmological Evolution Survey (COSMOS) field \citep{2007ApJS..172....1S}. For ASKAP, we use the simulated 2D visibility coverage from a 2 h tracking, pointing at RA 60$^{\circ}$, Dec. -30$^{\circ}$. For SKA1-MID, baseline the density is taken from the publicly available package {\tt Bao21cm}\footnote{\url{https://gitlab.com/radio-fisher/bao21cm}} at Dec. -30$^{\circ}$. Note that the tracking time is only used for simulating the visibility coverage to extract the baseline density. The observation time used for our forecasts can be found in Table \ref{tab:survey}. The derived baseline density is rescaled according to Eq. (\ref{eq:renormnu}).
As we only use the spherically averaged and normalized baseline density from the 2d visibility distribution, the direction of pointing and tracking time have very little impact on our forecasts as it negligibly changes the shape of the thermal noise power spectrum. More detailed investigations of the impact of survey strategies are beyond the scope of this work.

In this paper, we use finely gridded $k$-bins to calculate the noise, spanning across the minimum to maximum scale for each survey. In practice the $k$-bins will be coarser, but this choice is trivial and we report no significant difference in our results when choosing another set of $k$-bins.
\begin{figure}
\centering
\includegraphics[width=0.49\textwidth]{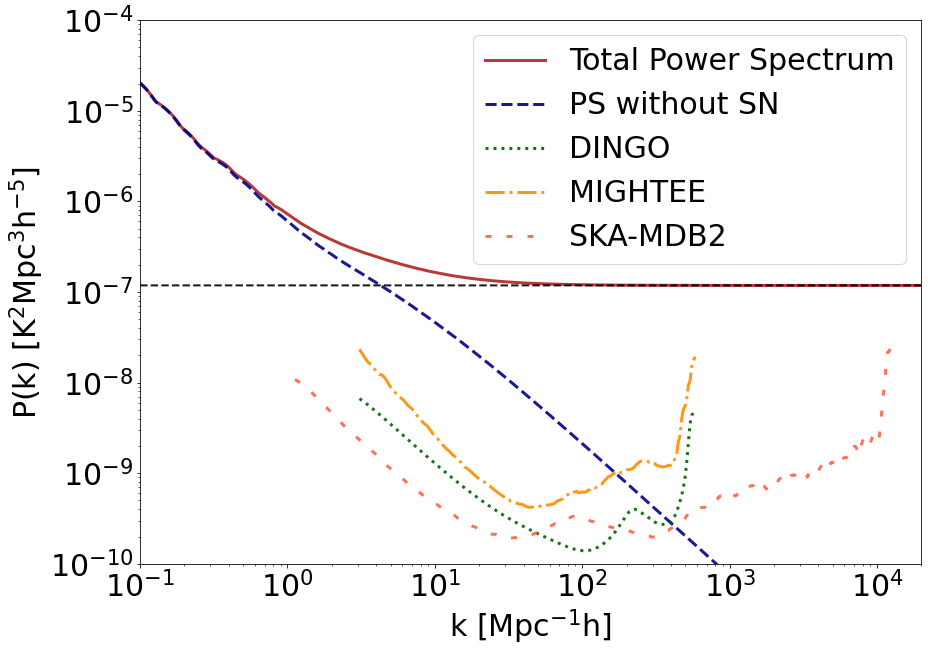}
\includegraphics[width=0.49\textwidth]{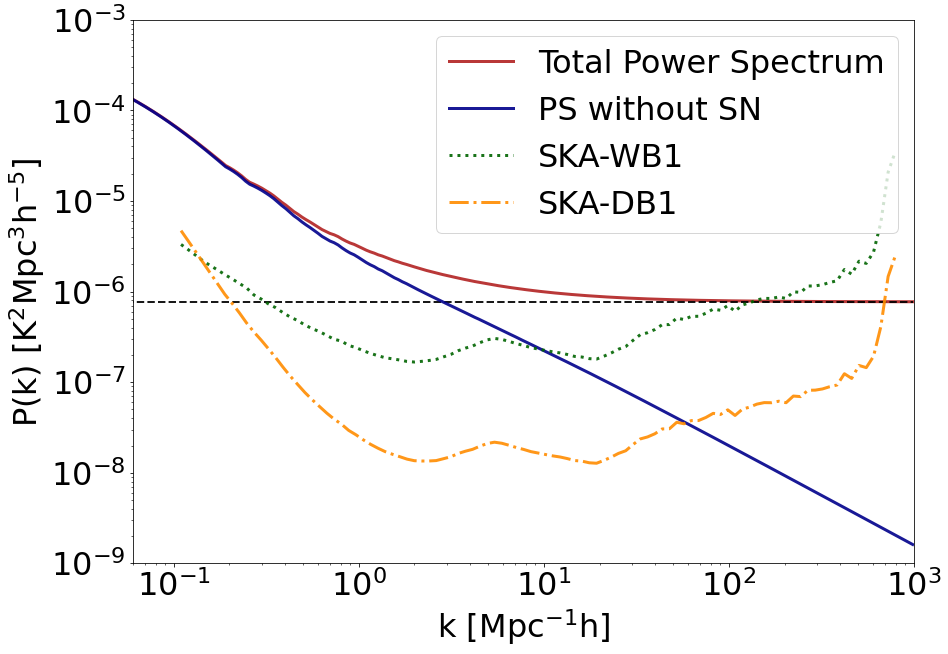}
\caption{{\it Upper panel:} the forecast for signal and noise of \hi\ temperature power spectrum at $z\sim0.1$, for MIGHTEE, DINGO and SKA-MDB2 surveys. {\it Lower panel:} the forecast for signal and noise of \hi\ temperature power spectrum at $z\sim1.0$, for SKA-WB1 and SKA-DB1 surveys. The total signal power spectrum is calculated using Eq. (\ref{eq:nonlinearp21}) with shot noise from simulation added on. The signal power spectrum without shot noise is also shown. The shape of the noise power spectrum is due to the non-monotonic baseline density $n(u)$ of the arrays.}
\label{fig:pknoise}
\end{figure}

The forecast for the detectability of the \hi\ power spectrum is presented in Fig. \ref{fig:pknoise}. The overall shape of the noise is determined by the baseline distribution of the array, where the smallest accessible scale is set by the longest baseline of the telescope. The amplitude of the noise of each experiment is a combined effect of system temperature, observation time, number of dishes and beams, and survey area, as all considered arrays have a similar dish size.

For $z\sim0.1$, the DINGO survey provides roughly two to four times better signal-to-noise ratio than MIGHTEE. Although the \hi\ power spectrum without shot noise at scales smaller than $k\sim 200 {\rm Mpc^{-1}}h$ is smaller than the noise, the additional small-scale $k$ bins facilitate constraining the shot noise to higher precision. The signal-to-noise ratio of the SKA-MDB2 is roughly two to three times higher on scales $k<10 {\rm Mpc^{-1}}h$, and due to longer baselines extends to much smaller scales.

For the SKA-WB1 survey at $z\sim1.0$, the \hi\ power spectrum without shot noise is detectable up to $k\sim 20 {\rm Mpc^{-1}}h$, and, similarly, on smaller scales the shot noise can be constrained. For the SKA-DB1 survey, the signal-to-noise ratio is improved by a factor of 10 compared to the wide survey. However, due to the smaller area, the deep survey is cosmic variance limited on the largest scales.

\section{Results}
\label{sec:results}
We perform the estimation of model parameters maximizing a Gaussian likelihood function with MCMC using the {\tt PYTHON} package {\tt EMCEE} \citep{2013PASP..125..306F}. 
Beside the parameters $[a_1^{\rm cen}, \beta_{\rm cen}, M_{\rm break}^{\rm cen},M_0^{\rm sat},\beta_{\rm sat},M_{\rm min}^{\rm sat},b]$ presented in Section \ref{sec:sim}, we further add the \hi\ shot noise $P_{\rm SN}$ (Section \ref{sec:SNform}) as a free parameter. We impose a large, flat prior on most parameters when running the chains as there are little to no observational constraints available. Since the  overall \hi\ density $\Omega_{\rm \hi}$ has been measured both by \hi\ galaxy surveys and damped Lyman $\alpha$ systems (DLAs; see discussion in Section \ref{sec:hihalo}), we impose a Gaussian prior on the derived parameter $\Omega_{\rm \hi}$, computed dynamically at each step. This helps in breaking the degeneracy between $\Omega_{\rm \hi}$ and $P_{\rm SN}$. Moreover, for $z\sim1$ we put strong lower bounds on parameters $M^{\rm cen}_{\rm break}$ and $M^{\rm sat}_{\rm min}$, due to the physical meaning of the parameters as discussed in Section \ref{sec:hihalo}. We list all the priors used in our fitting in Table \ref{tab:prior}.

We also impose a sanity check on \hi\ shot noise as Eq. (\ref{eq:PSN}) predicts a lower bound for \hi\ shot noise when setting $\sigma_{\rm \hi}=0$. Thus, in each step we calculate this lower bound $P_{\rm SN}^{\sigma_{\rm \hi}=0}$ accordingly, and compare with the free shot noise parameter. If the free shot noise parameter is smaller than this bound, the likelihood is set to be zero.

\begin{table*}
\centering
\begingroup
\setlength{\tabcolsep}{5pt} 
\begin{tabular}{cccccccccc}
  &  $a_1^{\rm cen}$ & $\beta_{\rm cen}$ & ${\rm log_{10}}\big[M^{\rm cen}_{\rm break}$ & ${\rm log_{10}}\big[M_0^{\rm sat}$ & $\beta_{\rm sat}$ & ${\rm log_{10}}\big[M^{\rm sat}_{\rm min}$ & $b$ & $P_{\rm SN}$ & $\Omega_{\rm \hi}$\\ 
  &  $[10^{-3}]$ &  & $/[M_\odot {h^{-1}}]\big]$ & $/[M_\odot {h^{-1}}]\big]$ &  & $/[M_\odot {h^{-1}}]\big]$ &  & ${\rm [Mpc^{-3}}h^3]$ & $[10^{-4}]$\\ \hline
$z\sim0.1$ &  $[0,{\textbf{4.66}},20]$ & $[0,{\textbf{0.85}},4]$ & $[5,{\textbf{10.66}},14]$ & $[4,{\textbf{9.51}},14]$ & $[0,{\textbf{0.81}},4]$ & $[7,{\textbf{12.00}},16]$ & $[0,{\textbf{2.26}},4]$ & $[P_{\rm SN}^{\rm min},{\textbf{47.45}},100]$ & $4.3^{+0.3}_{-0.3}$\\ 
$z\sim1.0$ &  $[0,{\textbf{3.00}},20]$ & $[0,{\textbf{0.43}},10]$ & $[9.5,{\textbf{11.86}},16]$ & $[1,{\textbf{8.58}},14]$ & $[0,{\textbf{1.10}},10]$ & $[10,{\textbf{11.40}},15]$ & $[0,{\textbf{2.45}},6]$ & $[P_{\rm SN}^{\rm min},{\textbf{61.00}},100]$ & $4.6^{+1.0}_{-1.0}$\\ \hline
\end{tabular}
\endgroup
\caption{Priors used in MCMC fitting. [] denotes flat prior, with lower bound, fiducial value and upper bound where $^+_-$ denotes Gaussian prior.
The lower bound for the shot noise parameter is defined as $P_{\rm SN}^{\rm min}\equiv{\rm max}\{10,P_{\rm SN}^{\sigma_{\rm \hi}=0}\}$, where
$P_{\rm SN}^{\sigma_{\rm \hi}=0}$ denotes the value computed for every step in the chain evaluating Eq. (\ref{eq:PSN}) at the values of the HOD parameters, considering $\sigma_{\rm \hi}=0$.}
\label{tab:prior}
\end{table*}
\begin{table*}
\centering
\begingroup
\setlength{\tabcolsep}{8pt} 
\renewcommand{\arraystretch}{1.2}
\begin{tabular}{ccccccccc}
  &  $a_1^{\rm cen}$ & $\beta_{\rm cen}$ & ${\rm log_{10}}\big[M^{\rm cen}_{\rm break}$ & ${\rm log_{10}}\big[M_0^{\rm sat}$ & $\beta_{\rm sat}$ & ${\rm log_{10}}\big[M^{\rm sat}_{\rm min}$ & $b$ & $P_{\rm SN}$\\ 
  &  $[10^{-3}]$ &  & $/[M_\odot {h^{-1}}]\big]$ & $/[M_\odot {h^{-1}}]\big]$ &  & $/[M_\odot {h^{-1}}]\big]$ &  & ${\rm [Mpc^{-3}}h^3]$ \\ \hline
$z\sim0.1$ & $4.66$ & $0.85$ & $10.66$ & $9.51$ & $0.81$ & $12.00$ & $2.26$ & $47.45$ \\ 
MIGHTEE & $4.92^{+1.82}_{-1.28}$ & $0.81^{+0.15}_{-0.14}$ & $10.67^{+0.18}_{-0.17}$ & $9.28^{+1.05}_{-0.69}$ & $0.80^{+0.09}_{-0.10}$ & $11.82^{+0.95}_{-0.75}$ & $2.26^{+0.03}_{-0.05}$ & $49.99^{+8.14}_{-6.59}$ \\ 
DINGO & $4.99^{+0.83}_{-0.73}$ & $0.81^{+0.09}_{-0.09}$ & $10.67^{+0.09}_{-0.10}$ & $9.19^{+0.54}_{-0.44}$ & $0.83^{+0.04}_{-0.05}$ & $11.69^{+0.53}_{-0.43}$ & $2.26^{+0.02}_{-0.02}$ & $49.82^{+6.81}_{-5.85}$ \\
SKA-MDB2 & $5.01^{+0.67}_{-0.57}$ & $0.82^{+0.06}_{-0.07}$ & $10.67^{+0.07}_{-0.06}$ & $9.41^{+0.34}_{-0.31}$ & $0.82^{+0.02}_{-0.03}$ & $11.91^{+0.33}_{-0.32}$ & $2.26^{+0.01}_{-0.01}$ & $47.66^{+1.28}_{-1.33}$ \\ \hline
$z\sim1.0$ &  $3.00$ & $0.43$ & $11.86$ & $8.58$ & $1.10$ & $11.40$ & $2.45$ & $61.00$ \\ 
SKA-WB1 &  $10.58^{+10.75}_{-7.87}$ & $0.63^{+1.04}_{-0.47}$ & $10.16^{+1.18}_{-0.49}$ & $9.63^{+1.07}_{-1.46}$ & $0.72^{+0.40}_{-0.36}$ & $11.72^{+1.02}_{-1.09}$ & $2.18^{+0.31}_{-0.47}$ & $64.16^{+13.30}_{-14.66}$ \\ 
SKA-DB1 &  $3.54^{+2.54}_{-1.32}$ & $0.38^{+0.16}_{-0.21}$ & $11.84^{+0.35}_{-0.34}$ & $8.56^{+1.18}_{-0.77}$ & $1.04^{+0.15}_{-0.13}$ & $11.35^{+0.96}_{-0.78}$ & $2.44^{+0.06}_{-0.08}$ & $59.44^{+6.38}_{-4.34}$ \\ 
\end{tabular}
\endgroup
\caption{The mean and 1$\sigma$ confidence interval of the marginalized model parameters given by the MCMC fit to the autopower spectrum. The $z\sim0.1$ and $z\sim1.0$ rows denote the fiducial values of the parameters at these redshifts. Each estimate is denoted with the survey name used for detectability forecasts in Section \ref{sec:det}.} 
\label{tab:pars}
\end{table*}

\subsection{Model Parameter Constraints}
\subsubsection{Model Parameter Constraints at $z\sim0.1$}
\begin{figure*}
\centering
\includegraphics[width=0.99\textwidth]{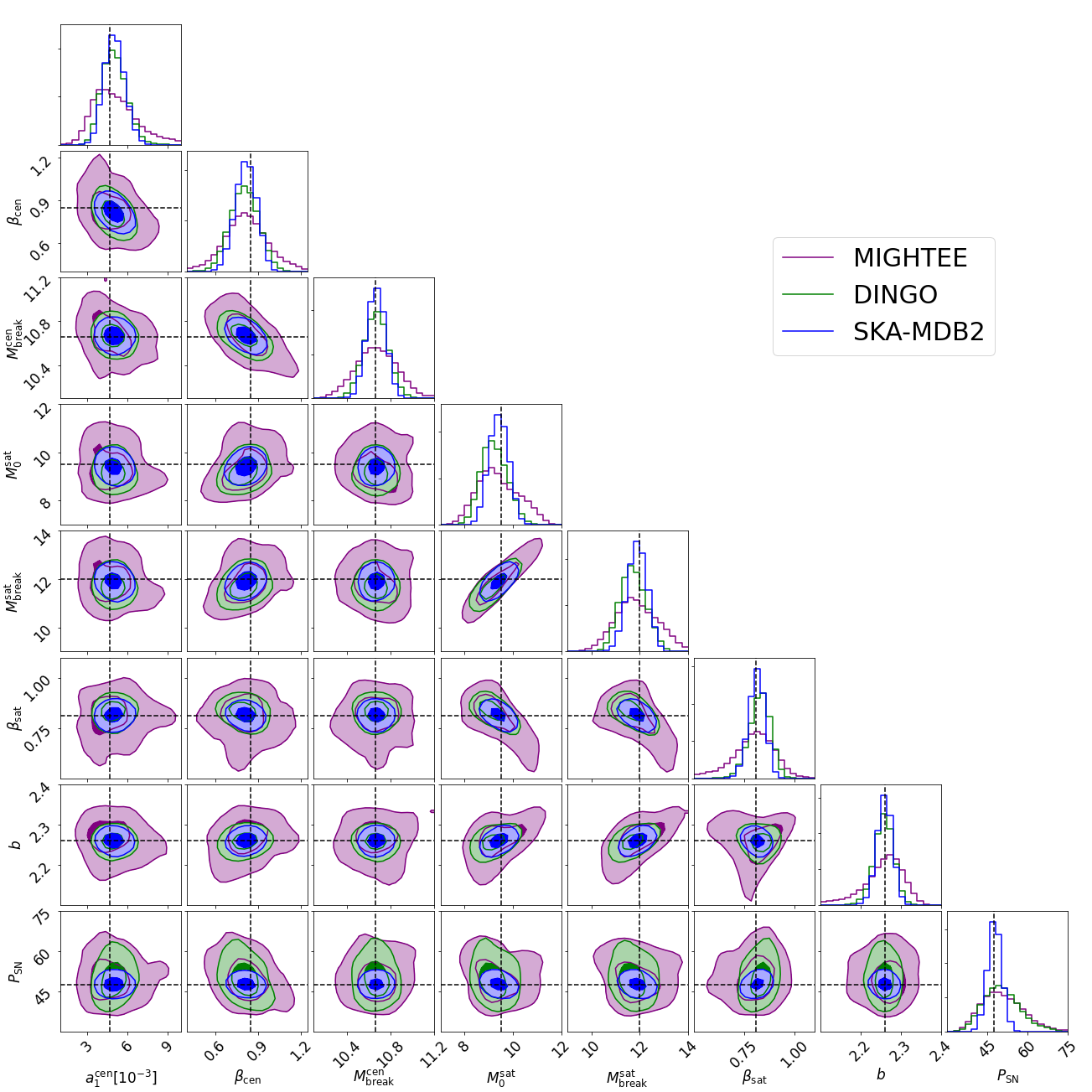}
\caption{The posterior distribution of the model parameters in 1$\sigma$ and 2$\sigma$ confidence levels for DINGO (in green), MIGHTEE (in purple), and SKA-MDB2 (in blue) at $z\sim0.1$. For better illustration, a Gaussian smooth kernel with a width of 1.0 is applied to the data array. The dashed lines represent fiducial values.}
\label{fig:parcorner}
\end{figure*}

We present the mean and 1$\sigma$ confidence interval for the marginalized model parameters in Table \ref{tab:pars} and the posterior distributions in Fig. \ref{fig:parcorner} for parameter estimation at $z\sim0.1$. 
For all surveys considered (MIGHTEE, DINGO, and SKA), the fiducial values of each parameter are within the 1$\sigma$ confidence interval of our estimation.

The parameters $a_1$, $M_0$, $\beta_{\rm cen,sat}$ are poorly constrained due to the degeneracies discussed in Section \ref{sec:hihalo}. Besides that, $M^{\rm break}_{\rm sat}$ only serves as a small mass cut-off, and is relatively poorly constrained. 

Because of the accurate measurement of the power spectrum at small scales, $b$ is well constrained. Overall, error bars at the 10\% level can be achieved in the near future for MIGHTEE and DINGO, while SKA-MDB2 offers roughly a factor of 2 improvement.

Although there is a large signal-to-noise ratio for power spectrum measurements on small scales, the \hi\ shot noise measurement in length unit is slightly biased towards large values for MIGHTEE and DINGO, due to the fact that $\Omega_{\rm HI}$ and \hi\ shot noise have exact anticorrelation.

From the posterior distribution, we can further compare the constraining power of the different surveys. Note that the posterior of MIGHTEE is non-Gaussian due to the relatively small signal-to-noise ratio. This is massively improved by DINGO, as DINGO measures the power spectrum with extreme precision at $k\sim 100\,{\rm Mpc^{-1}}{h}$. Therefore, we conclude that to resolve the problem of degeneracies of model parameters, an accurate measurement (a signal-to-noise ratio around 10) of the power spectrum up to $k\sim 100\,{\rm Mpc^{-1}}{h}$ is needed. Comparing DINGO with SKA-MDB2, we further conclude that adding measurements on large scales ($k\sim 1\,{\rm Mpc^{-1}}{h}$) can further improve the constraints.

\subsubsection{Model Parameter Constraints at $z\sim1.0$}

\begin{figure*}
\centering
\includegraphics[width=0.99\textwidth]{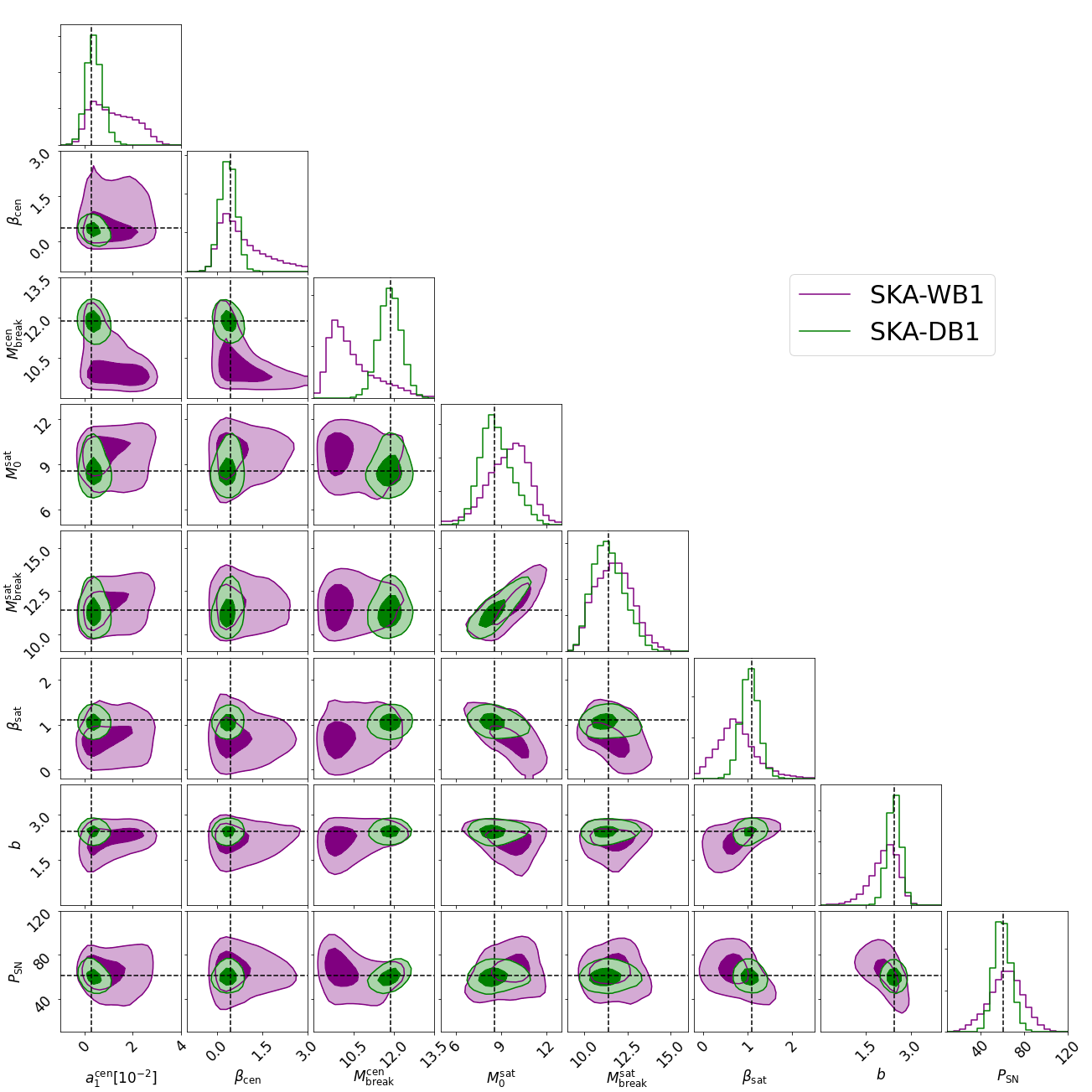}
\caption{The posterior distribution of the model parameters in 1$\sigma$ and 2$\sigma$ confidence levels for SKA-WDB1 (in purple) and SKA-DB1 (in green) at $z\sim1.0$. For better illustration, a Gaussian smooth kernel with a width of 1.0 is applied to the data array. The dashed lines represent fiducial values.}
\label{fig:parcorner2}
\end{figure*}

For the shallow SKA-WB1 survey at $z\sim1.0$, the \hi\ power spectrum can only be measured up to $k\sim 10 {\rm Mpc^{-1}}{h}$ unlike at $z\sim0.1$, where additionally a small prior on $\Omega_{\rm \hi}$ helps break parameter degeneracies. This leads to a poor fitting performance for SKA-WB1, as seen in the results presented in Table \ref{tab:pars} and Fig. \ref{fig:parcorner2}. 

At this higher redshift, a shallow survey can measure the \hi\ power spectrum only up to $k\sim 10 {\rm Mpc^{-1}}{h}$. Moreover, our prior on $\Omega_{\rm \hi}$ is less tight and thus less effective in breaking parameter degeneracies. These reasons lead to a poor fitting performance for SKA-WB1.

Most noticeably, the tails of $a_1^{\rm cen}$, $\beta_{\rm cen}$ and $M_{\rm break}^{\rm cen}$ in the histogram suggest the existence of a local minimum. The posterior distribution clearly shows the existence of two minima: the fiducial parameter set with higher $M_{\rm break}^{\rm cen}$ and much smaller $a_1^{\rm cen}$, $\beta_{\rm cen}$; and the false minimum with a very low $M^{\rm break}_{\rm cen}$, which leads to the fact that $a_1^{\rm cen}$, $\beta_{\rm cen}$ can be almost arbitrarily large.

The improved signal-to-noise ratio from the deep SKA-DB1 survey solves the occurrence of the local minimum. The trend we see in $z\sim0.1$, namely the overestimation of $a_1^{\rm cen}$ and its impact on other parameters, is still present. Again, all fiducial values are within the 1$\sigma$ confidence interval of our estimation.

\subsection{Transforming to Physical Parameters}
\label{subsec:phy}
\begin{table*}
\centering
\begingroup
\setlength{\tabcolsep}{6pt} 
\renewcommand{\arraystretch}{1.2}
\begin{tabular}{cccccccc|c}
 &  $\Omega_{\rm \hi}$   & $b_{\rm \hi}^0$ & $b_{\rm \hi}^{1 {\rm Mpc^{-1}}{h}}$ & $b_{\rm \hi}^{10 {\rm Mpc^{-1}}{h}}$ & $\tilde{M}_{\rm \hi}^{10^{11}M_\odot{h^{-1}}}$  & $\tilde{M}_{\rm \hi}^{10^{12}M_\odot{h^{-1}}}$  & $\tilde{M}_{\rm \hi}^{10^{13}M_\odot{h^{-1}}}$ & $\sigma_{\rm \hi}$  \\ 
  &  [10$^{-4}$]   &  &  &  & [10$^{-12}$]  & [10$^{-13}$]  & [10$^{-15}$] &  \\ \hline
$z\sim0.1$ & 4.30 & 0.94 & 0.72 & 0.75 & 1.39 & 1.75 & 5.41 & /\\ 
MIGHTEE & $4.19_{-0.30}^{+0.31}$ & $0.94_{-0.02}^{+0.02}$ & $0.74_{-0.04}^{+0.05}$ & $0.77_{-0.07}^{+0.09}$ & $1.37_{-0.19}^{+0.19}$ & $1.66_{-0.22}^{+0.21}$ & $5.47_{-0.99}^{+0.80}$ & $1.19^{+0.56}_{-0.41}$\\
DINGO & $4.20_{-0.26}^{+0.27}$ & $0.94_{-0.01}^{+0.01}$ & $0.74_{-0.04}^{+0.04}$ & $0.77_{-0.05}^{+0.06}$ & $1.38_{-0.10}^{+0.09}$ & $1.66_{-0.13}^{+0.15}$ & $5.43_{-0.37}^{+0.38}$ & $1.25^{+0.34}_{-0.30}$\\
SKA-MDB2 & $4.30_{-0.06}^{+0.06}$ & $0.94_{-0.003}^{+0.004}$ & $0.73_{-0.01}^{+0.01}$ & $0.75_{-0.02}^{+0.02}$ & $1.40_{-0.06}^{+0.07}$ & $1.69_{-0.09}^{+0.08}$ & $5.43_{-0.16}^{+0.16}$ & $1.13_{-0.14}^{+0.16}$ \\\hline
$z\sim1.0$ & 4.61 & 1.34 & 1.37 & 1.30 & 1.22 & 1.84 & 6.23 & / \\
SKA-WB1 & $4.63_{-0.33}^{+0.43}$ & $1.34_{-0.11}^{+0.10}$ & $1.45_{-0.20}^{+0.18}$ & $1.73_{-0.59}^{+1.10}$ & $1.44_{-0.73}^{+0.77}$ & $1.26_{-0.81}^{+0.69}$ & $8.50_{-3.78}^{+2.62}$ & $1.53_{-0.85}^{+2.17}$ \\
SKA-DB1 & $4.68_{-0.23}^{+0.17}$ & $1.32_{-0.05}^{+0.07}$ & $1.36_{-0.05}^{+0.07}$ & $1.32_{-0.11}^{+0.14}$ & $1.34_{-0.23}^{+0.24}$ & $1.69_{-0.35}^{+0.22}$ & $6.45_{-1.67}^{+1.81}$ & $1.27_{-0.46}^{+0.73}$ \\

\end{tabular}
\endgroup
\caption{The mean and  1$\sigma$ confidence interval of marginalised physical parameters derived from the MCMC fit to the auto-power spectrum, as discussed in Section \ref{subsec:phy}. The $z\sim0.1$ and $z\sim1.0$ rows represent the fiducial values derived from the simulation at these redshifts. The last column for $\sigma_{\rm \hi}$ is introduced in Section \ref{sec:SN}. }
\label{tab:pars2}
\end{table*}

The parameters presented above are particular to our model and cannot be easily compared with other results and notations. Thus, we transform them to obtain new parameters with clearer physical meaning.

The measured power spectrum can be described by the \hi\ density $\Omega_{\rm \hi}$ and the \hi\ bias at various scales \{$b_{\rm \hi}^{k_i}$\}. We also seek to interpret the HOD in a parameterisation-independent fashion. Therefore we transform the model parameter set into:
\begin{equation}
\begin{split}
\Big\{ \Omega_{\rm \hi}, b_{\rm \hi}^0, b_{\rm \hi}^{1 {\rm Mpc^{-1}}{h}}&, b_{\rm \hi}^{10 {\rm Mpc^{-1}}{h}}, \\
\tilde{M}_{\rm \hi}^{10^{11}M_\odot{h^{-1}}}, &\tilde{M}_{\rm \hi}^{10^{12}M_\odot {h^{-1}}}, \tilde{M}_{\rm \hi}^{10^{13}M_\odot{ h^{-1}}}\Big\}
\end{split}
\end{equation} 
where $b_{\rm \hi}^0$ is the \hi\ bias at large scales (i.e. $k \to 0\: {\rm Mpc^{-1}}{h}$) and
the dimensionless $\tilde{M}_{\rm \hi}^m$ is a {renormalized} \hi\ HOD defined as
\begin{equation}
\tilde{M}_{\rm \hi}^m = n(m) \langle M_{\rm \hi}(m) \rangle/\bar{\rho}_{\rm \hi}.
\end{equation}
This quantifies the contribution of different halo masses to the total \hi\ density. We show our results for $10^{11},10^{12},10^{13} M_\odot {h^{-1}}$ as most \hi\ in our simulation resides in halos in this mass range.

The degeneracy between $\Omega_{\rm \hi}$ and $b_{\rm \hi}$ breaks down due to small-scale information and scale dependency of the \hi\ bias modelled with {\tt halomod} that allows us to constrain them separately. Combined with redshift space distortions (e.g. \citealt{2015PhRvD..92j3516O}), this could be critical for future observations to isolate cosmological parameters.

We present the mean and 1$\sigma$ confidence interval of the marginalised physical parameters in Table \ref{tab:pars2}.
\begin{figure*}
\centering
\includegraphics[width=0.99\textwidth]{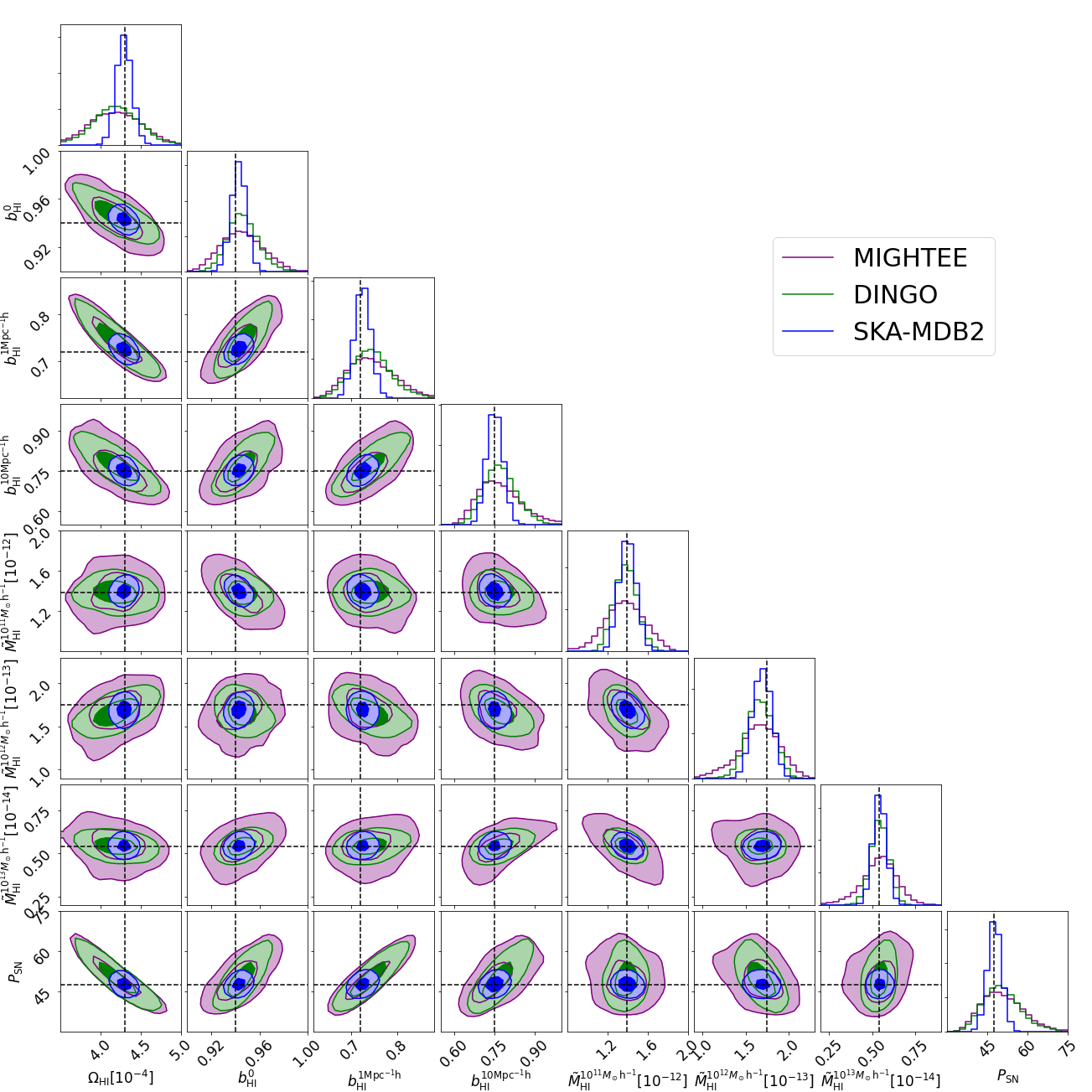}
\caption{The posterior distribution of the physical parameters in 1$\sigma$ and 2$\sigma$ confidence levels for MIGHTEE (in purple), DINGO (in green), and SKA-MDB2 (in blue) at $z\sim0.1$. For better illustration, a Gaussian smooth kernel with a width of 1.0 is applied to the data array. The dashed lines represent fiducial values.}
\label{fig:parcorner3}
\end{figure*}
\begin{figure*}
\centering
\includegraphics[width=0.99\textwidth]{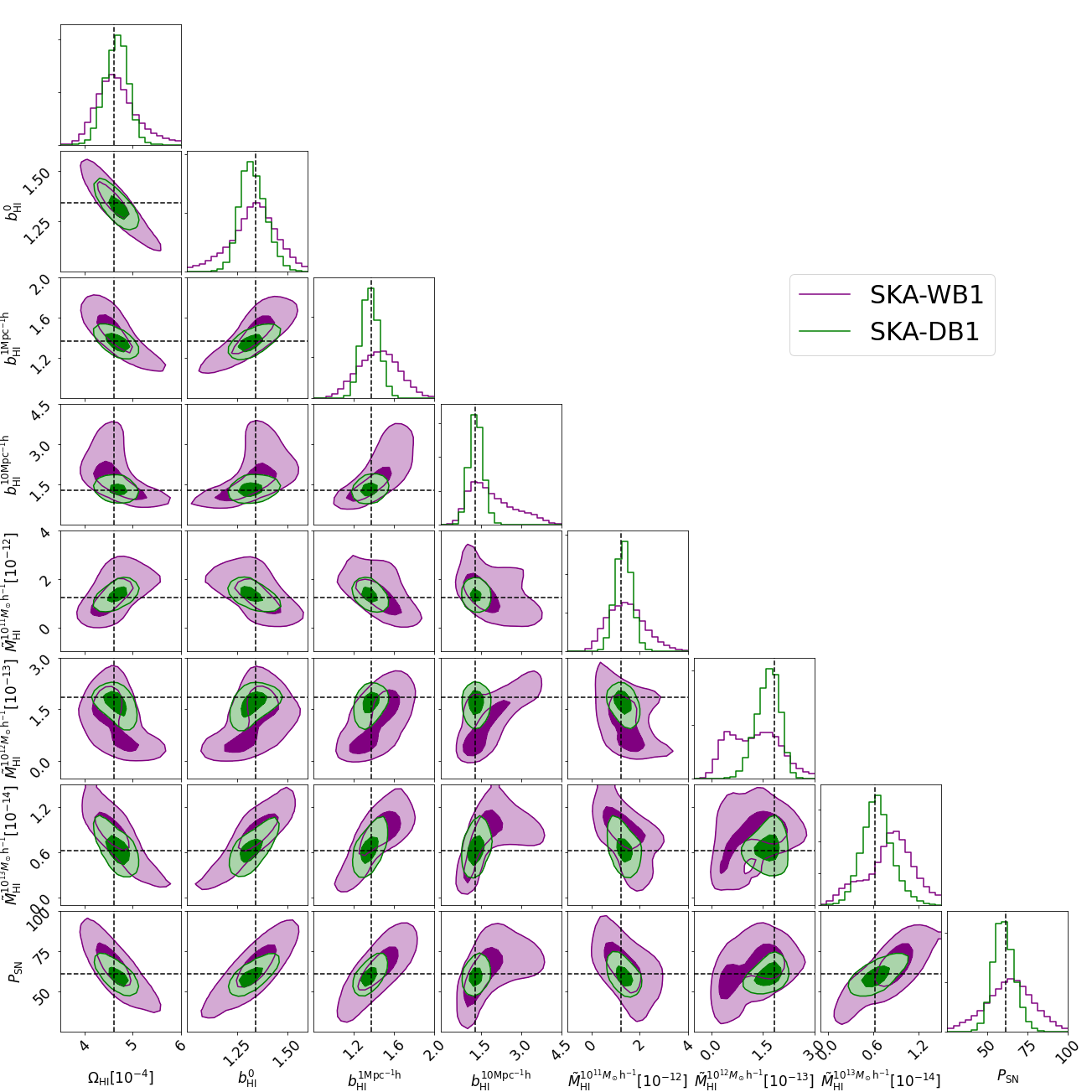}
\caption{The posterior distribution of the physical parameters in 1$\sigma$ and 2$\sigma$ confidence levels for SKA-WB1 (in purple) and SKA-DB2 (in green) at $z\sim1.0$. For better illustration, a Gaussian smooth kernel with a width of 1.0 is applied to the data array. The dashed lines represent fiducial values.}
\label{fig:parcorner4}
\end{figure*}
\subsubsection{Physical Parameter Constraints at $z\sim0.1$}
For $z\sim0.1$, we obtain a good estimation of physical parameters as we can see from Table \ref{tab:pars2} and Fig. \ref{fig:parcorner3}. 
For all surveys, constraints on $\Omega_{\rm \hi}$ comparable or below $3\times10^{-5}$ can be achieved. Considering the fact that we impose a $4.30_{-0.30}^{+0.30}$ Gaussian prior to control parameter degeneracy, the constraining power mainly comes from the existing measurement as prior instead of intensity mapping. Nevertheless, it provides a proof of concept to use intensity mapping to measure $\Omega_{\rm \hi}$, and can be applied to higher redshifts as we discuss later. 
Tight constraints on $b_{\rm \hi}$ are simultaneously obtained. For example, for linear bias $b_{\rm \hi}^0$, we forecast an estimation of $b_{\rm \hi}^0=0.94^{+0.02}_{-0.02}$ for MIGHTEE and $b_{\rm \hi}^0=0.94^{+0.01}_{-0.01}$ for DINGO. 

Comparing the results from MIGHTEE and DINGO, we note that the improvements coming from better signal-to-noise ratio mainly benefit the small-scale measurement, namely $b_{\rm \hi}^{10{\rm Mpc^{-1}}{h}}$ in our parameter set. As a result of this, renormalized \hi\ HODs are better constrained.

From Fig. \ref{fig:parcorner3}, one can see the expected anticorrelation between $P_{\rm SN}-\Omega_{\rm \hi}$ and $b_{\rm \hi}^{k}-\Omega_{\rm \hi}$. More interestingly, we note that ${M}_{\rm \hi}^{10^{11}M_\odot{ h^{-1}}}$ is anticorrelated with $b_{\rm \hi}^0$, while mostly uncorrelated with \hi\ bias at higher $k$. ${M}_{\rm \hi}^{10^{12}M_\odot{h^{-1}}}$ is anti-correlated with $b_{\rm \hi}^{1{\rm Mpc^{-1}}{h}}$ and $b_{\rm \hi}^{10{\rm Mpc^{-1}}{h}}$, while mostly uncorrelated with $b_{\rm \hi}^0$. Lastly, ${M}_{\rm \hi}^{10^{13}M_\odot{h^{-1}}}$ is slightly correlated with $b_{\rm \hi}$. These correlations provide the insights into how halos of different mass affect \hi\ clustering.

Similar to the model parameters, we find that measurements of the power spectrum up to $k\sim100\,{\rm Mpc^{-1}}{h}$ are needed for physical parameter posteriors to fully converge. The extra measurements at $k\sim1\,{\rm Mpc^{-1}}{h}$ of SKA-MDB2 further improve the estimation of physical parameters.

\subsubsection{Physical Parameter Constraints at $z\sim1.0$}
For SKA-WB1, we can see in Table \ref{tab:pars2} and Fig. \ref{fig:parcorner4} that although the constraints on model parameters are poor, the overall amplitude of the spectrum $\Omega_{\rm \hi}$ and $b_{\rm \hi}^0$ are within 10\%. The large error bar on $b_{\rm \hi}^{10{\rm Mpc^{-1}}{h}}$ shows that the poor measurement on small scales produces the false minimum in the posterior, which can be isolated by looking at physical parameter space instead of model parameter space. As a result, the \hi\ HOD at various mass scales is poorly constrained. Most noticeably, one can see the bimodal distribution of \hi\ HOD at $m=10^{12}{M_\odot }{h^{-1}}$, the mass scale around which most \hi\ resides. 

Even though the SKA-WB1 can only constrain the \hi\ power spectrum up to a few ${\rm Mpc^{-1}}{h}$, the measurement of shot noise goes down to much smaller scales. Combined with the fact that $\Omega_{\rm \hi}$ is accurately inferred, the shot noise parameter $P_{\rm SN}$ is also relatively well constrained.

The constraints for the deep survey SKA-DB1 on \hi\ density and linear bias improve to $\Omega_{\rm \hi}=4.68_{-0.23}^{+0.17}$ and $b_{\rm \hi}^0 = 1.32_{-0.05}^{+0.07}$. The improvement on $b_{\rm \hi}^0$ is about a factor of 2, whereas for $b_{\rm \hi}^{10{\rm Mpc^{-1}}{h}}$, the improvement is about a factor of 10, which comes from the measurement on small scales. As a result of this, the estimation of \hi\ HOD is much more accurate and the bimodal distribution disappears. This again emphasizes the fact that in order to understand the \hi\ HOD, a measurement of small-scale \hi\ clustering is necessary.

\section{Extracting Information from Shot Noise}
\label{sec:SN}
In the previous section, we discussed the constraints obtained from the MCMC, which includes a measurement of \hi\ shot noise. As we are interested in extracting extra information from it, in this section we consider two possible approaches of doing so. In Section \ref{subsec:sigmahi}, shot noise is used to estimate the \hi-HOD scatter $\sigma_{\rm \hi}$. In Section \ref{subsec:HIMF}, we investigate the possible estimation of the \hi\ignorespaces MF from the shot noise measurement.
\subsection{Constraining \althi-HOD scatter}
\label{subsec:sigmahi}
\begin{figure}
\centering
\includegraphics[width=0.49\textwidth]{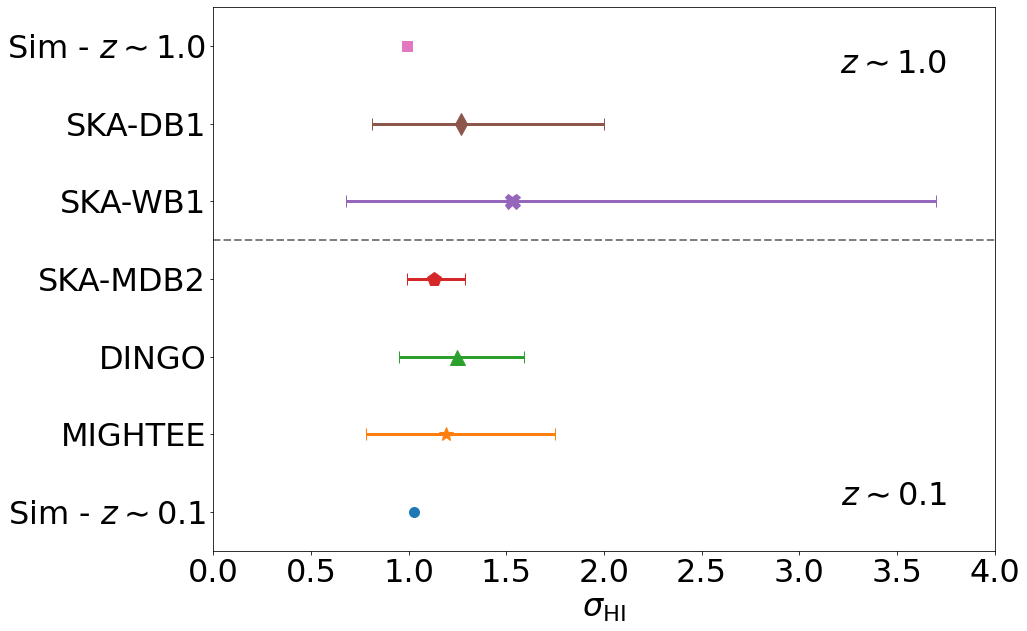}
\caption{The \hi-HOD scatter $\sigma_{\rm \hi}$ as an integrated average from our simulation and the estimated $\sigma_{\rm \hi}$ by treating it as a single constant in Eq.~(\ref{eq:PSN}). The dots represent the mean value and the error bars denote the 1$\sigma$ confidence intervals of the estimations.}
\label{fig:sigmahi}
\end{figure}
Using the shot noise expression of Eq.~(\ref{eq:PSN}), we treat the scatter of \hi-HOD, $\sigma_{\rm \hi}^{\rm cen,sat}(m)$, as a single constant $\sigma_{\rm \hi}$ and use the halo model parameters in our likelihood estimation to calculate $\sigma_{\rm \hi}$. 
Since in most halo mass ranges central \hi\ dominates, the single constant agrees closely with $\sigma_{\rm \hi}^{\rm cen}(m)$.
In each step, we plug the free parameter $P_{\rm SN}$ and HOD parameters into Eq.~(\ref{eq:PSN}) to calculate a value of $\sigma_{\rm \hi}$, and we present our results in Table \ref{tab:pars2}.

Since we treat $\sigma_{\rm \hi}$ as a single constant, we effectively measure the integrated average of the \hi-HOD scatter, as shown in Fig. \ref{fig:sigmahi}. We report that the shot noise contribution from \hi\ in central galaxies dominates and comprises roughly 99\% of the total shot noise. Therefore, in our case we effectively measure $\sigma_{\rm \hi}^{\rm cen}$ as an integrated average.

At $z\sim0.1$, MIGHTEE can constrain $\sigma_{\rm \hi}$ up to 50\% precision, whereas DINGO can improve it to 25\% and SKA-MDB2 further to around 10\%. At $z\sim1.0$, SKA-DB1 can constrain this scatter to around 40\%. 

We emphasize that the constraints listed here are an optimistic estimation since we assume the knowledge of the galaxy HOD. Realistically one needs a complimentary galaxy survey, most preferably matching the survey we assume for intensity mapping, to obtain an estimation of the galaxy HOD. In our simulation, the $\sigma_{\rm \hi}$ is calculated in a way that includes {\it all} galaxies. In practice, the HOD of galaxies, estimated from a sample of galaxies selected from a sensitivity-limited survey, can be used.

\subsection{Shot Noise in terms of \althi\ Mass Function}
\label{subsec:HIMF}
In this section, we use the measured shot noise to constrain the \hi\ignorespaces MF as discussed in Section \ref{sec:SNform}. In order to impose a valid constraint on the \hi\ignorespaces MF parametrized by a Schechter function, one needs estimations of the \hi\ source density $n_{\rm g}$, \hi\ density $\Omega_{\rm \hi}$ and $P_{\rm SN}$. Conveniently, from the power spectrum we already have an estimation of $\Omega_{\rm \hi}$ and $P_{\rm SN}$, with only $n_{\rm g}$ missing. To obtain an estimation of the number density of \hi\ galaxies, once again a complimentary galaxy survey is needed. However, we do not require an ultradeep \hi\ galaxy survey for this. Only galaxies with \hi\ mass larger than a certain mass threshold $M_{\rm \hi,min}$ contribute to the overall \hi\ density and \hi\ shot noise. In our simulation, considering existing measurements reported in eg. \cite{2010ApJ...723.1359M}, this mass threshold is $M_{\rm \hi,min}\sim 10^7 { M_\odot h^{-1}}$ for less than 1\% deviation and $M_{\rm \hi,min}\sim 10^8 {M_\odot h^{-1}}$ for less than 5\% deviation. This is already above the minimum mass of \hi\ galaxy sample reported in \cite{2010ApJ...723.1359M}. Additionally, we do not need to know the exact \hi\ mass of these galaxies, but only their number density. If the number density of galaxies is given, we can estimate the threshold for \hi\ mass in this sample and use it as the lower bound of the integration for Eqs. (\ref{eq:ng}), (\ref{eq:omegahimf}), and (\ref{eq:himf}). In our case, since we have good knowledge of \hi\ galaxies from the simulation, we start from $10^6 M_\odot{h^{-1}}$. This selection is trivial and will not affect the result as long as the number density of \hi\ galaxies and the mass threshold are consistent with each other.

The $\Omega_{\rm \hi}$ and $P_{\rm SN}$ are obtained from MCMC fitting, and used to derive the constraints on the \hi\ignorespaces MF. Here we present the results, for both $z\sim0.1$ and $z\sim1.0$, in Table \ref{tab:schechter}. This approach can be generalised and applied to higher redshifts.

\begin{table}
\centering
\begingroup
\setlength{\tabcolsep}{4pt} 
\renewcommand{\arraystretch}{1.2}
\begin{tabular}{cccc}
 & $\phi_*/10^{-3}$ & ${\rm log}_{10} \big[M_*/M_\odot\big]$ & $\alpha$\\
  & $[h_{70}^3{\rm Mpc^{-3}dex^{-1}}]$ & $+2{\rm log}_{10}[h_{70}]$&\\\hline
J18 & $4.5^{+0.2}_{-0.2}$ & $9.94^{+0.01}_{-0.01}$ & $-1.25^{+0.02}_{-0.02}$\\
MIGHTEE & $4.05^{+0.96}_{-0.73}$ & $10.05^{+0.08}_{-0.09}$ & $-1.33^{+0.02}_{-0.02}$\\
DINGO & $4.04^{+0.82}_{-0.63}$ & $10.06^{+0.07}_{-0.07}$ & $-1.33^{+0.02}_{-0.02}$\\
SKA-MDB2 & $4.18^{+0.16}_{-0.15}$ & $10.06^{+0.01}_{-0.02}$ & $-1.32^{+0.004}_{-0.004}$\\\hline
SKA-WB1 & $3.64^{+2.66}_{-1.07}$ & $10.15^{+0.14}_{-0.21}$ & $-1.33^{+0.06}_{-0.03}$\\
SKA-DB1 & $3.13^{+0.38}_{-0.36}$ & $10.20^{+0.05}_{-0.05}$ & $-1.34^{+0.01}_{-0.01}$\\
\end{tabular}
\endgroup
\caption{Marginalized parameter likelihood of 1$\sigma$ confidence interval derived from the MCMC fit to the shot noise according to Eq. (\ref{eq:himf}), with the constraints obtained from local \hi\ galaxy survey in \protect\cite{2018MNRAS.477....2J} ('J18') for comparison. Only the statistical error of J18 is shown since our work does not forecast systematics.}
\label{tab:schechter}
\end{table}

The Schechter parameters for the \hi\ignorespaces MF in the local Universe have been measured using \hi\ galaxy surveys \citep{2005MNRAS.359L..30Z,2010ApJ...723.1359M, 2018MNRAS.477....2J}. Thus as a comparison, we present the results of \cite{2018MNRAS.477....2J} together with our parameter likelihood in Table \ref{tab:schechter}. We emphasize that only the size of the error bars is comparable, since our simulation is in higher redshift than the measured \hi\ mass function. For $z\sim0.1$, MIGHTEE can constrain $\phi_*$ with a precision of around 10\%. For $M_*$ and $\alpha$ the precision can be around 2\%. The results from DINGO are slightly improved compared to MIGHTEE. To reach the precision of previous \hi\ galaxy surveys though, it will require a survey matching the scales and precision of SKA-MDB2. For $z\sim1.0$, we report an estimation of $\phi_*$ with 10\% precision, $M_*$ with 5\% precision and $\alpha$ with 1\% precision with SKA-DB1.

This formalism can serve as a check for systematics for \hi\ galaxy surveys and opens a new window to measure the \hi\ignorespaces MF at higher redshifts. 

We emphasize that the methods proposed are proofs of concept, and will be further investigated in future work towards validity.

\section{Conclusion}
\label{sec:conclusion}
\begin{figure}
\centering
\includegraphics[width=0.49\textwidth]{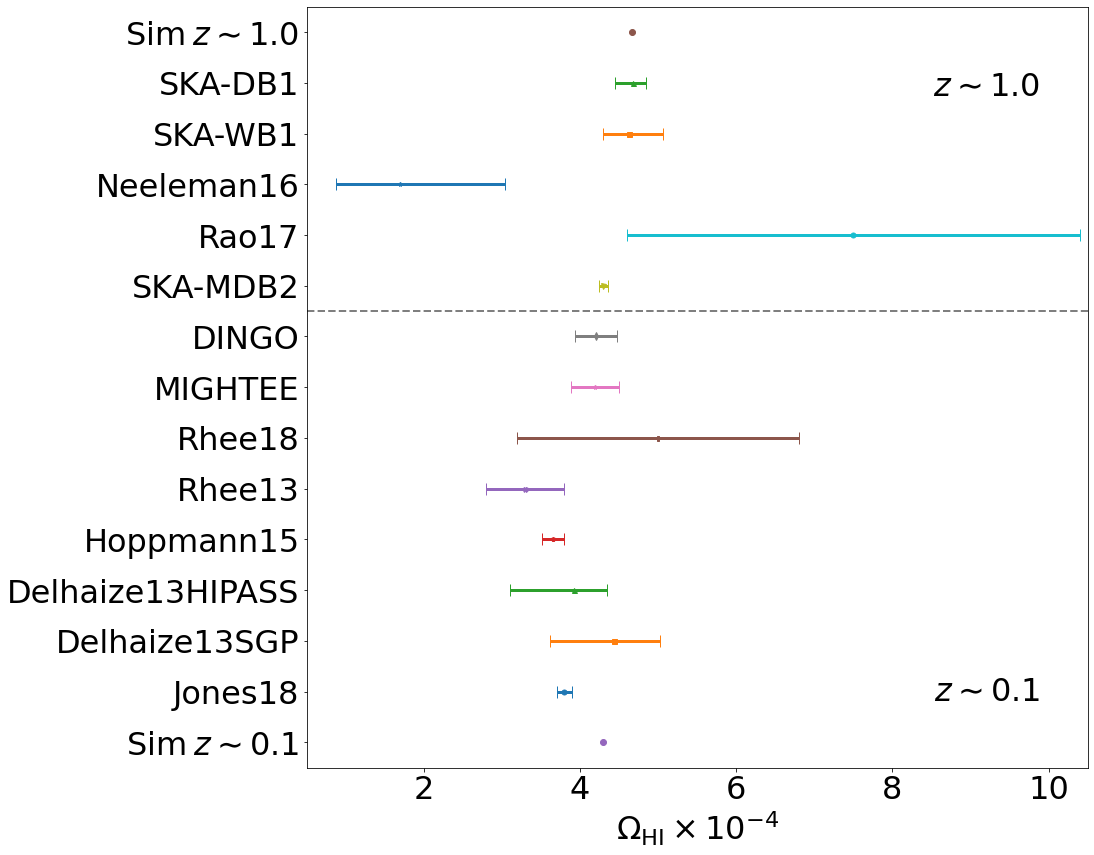}
\caption{Forecasts on the precision of $\Omega_{\rm \hi}$ measurements with comparison to existing data. The top panel includes our results at $z\sim0.1$, with the results from \hi\ galaxy survey of ALFALFA \protect\citep{2018MNRAS.477....2J}, Parkes \protect\citep{2013MNRAS.433.1398D}, AUDS \protect\citep{2015MNRAS.452.3726H}, WSRT \protect\citep{2013MNRAS.435.2693R}, GMRT \protect\citep{2018MNRAS.473.1879R} for $z<0.3$. The bottom panel includes our results at $z\sim1.0$, with results from DLA using SDSS \protect\citep{2017MNRAS.471.3428R,2016ApJ...818..113N} around $z\sim1.0$.}
\label{fig:omegahi}
\end{figure}
In this paper, we have explored how \hi\ astrophysics from future \hi\ intensity mapping surveys with interferometric arrays can be inferred with the halo model.
We model the \hi\ HOD and density profile according to the simulation work of \cite{2020MNRAS.493.5434S}, assuming survey strategies corresponding to MIGHTEE, DINGO, SKA-MDB2, SKA-WB1 and SKA-DB1 surveys, and use MCMC to  forecast the constraining power on our model parameters. The main results of our survey predictions for the \hi\ halo model are as follows.
\begin{itemize}
    \item 
    Using the halo model, we can reconstruct an accurate description of \hi\ clustering down to small scales $k>10{\rm Mpc^{-1}}{h}$. We use the \hi\ HOD and \hi\ satellite density profile in our simulations to obtain parametrized relations that can be used in our inference tool. We explore parameter degeneracies and find that eight parameters are required to universally describe the \hi\ halo model and the predicted \hi\ power spectrum.
    \item
    For $z\sim0.1$, we find that due to parameter degeneracy, the improved signal-to-noise ratio from DINGO compared to MIGHTEE on scales $k\sim 100{\rm Mpc^{-1}}{h}$ is necessary to fully distinguish different parameters such as $a_1^{\rm cen}$ and $\beta_{\rm cen}$. Switching to the physical parameter space shows that constraints for $\Omega_{\rm \hi}$ up to 6\% level precision and for $b_{\rm \hi}^0$ up to 1\% level precision can be achieved. 
    SKA-MDB2 improves the estimation by another factor of three, due to the improved sensitivity as well as covering larger scales $k\sim 1{\rm Mpc^{-1}}{h}$.
    \item
    For $z\sim1.0$, the wide SKA-WB1 survey does not provide enough information on small scales to fully determine the parameter set. Looking at physical parameter space, we find that it nevertheless provides a solid estimate of the overall \hi\ density and \hi\ shot noise, due to measurement of the power spectrum on large scales. A deep survey using SKA can improve the constraints on $\Omega_{\rm \hi}$ and $b_{\rm \hi}^0$ up to 5\% level precision.
\end{itemize}

The comparison of the constraining power of different surveys comes from planned survey specifications as mentioned in Section \ref{sec:det}.
We present a comparison of $\Omega_{\rm \hi}$ constraints from these surveys with existing measurements in Fig. \ref{fig:omegahi}.

We also examined the shot noise extensively in simulation and analytical formalism, resulting in the following conclusions.
\begin{itemize}
    \item
    We derive a new analytical formula to calculate \hi\ shot noise in the context of the halo model based on the \hi\ mass scatter within the discrete sources, Eq. (\ref{eq:PSN}). We find that the analytical shot noise matches the simulation well, opening a new way of associating \hi\ shot noise with \hi-HOD scatter in galaxies.
    \item
    We forecast that future surveys can constrain the \hi-HOD scatter as an integrated average, which can be vital when confronting simulations with observations. We predict that a future SKA survey can constrain this parameter within 10\% for $z\sim0.1$. 
    \item
    We also explore how to relate the \hi\ mass function to shot noise measurements. With the \hi\ density already obtained from fitting the power spectrum, only the number density of \hi\ galaxies for $M_{\rm \hi}>10^7 M_\odot {h^{-1}}$ is required to constrain the shot noise. We forecast that future SKA surveys can constrain the \hi\ mass function to the precision level of the latest local Universe \hi\ galaxy survey (e.g. \citealt{2018MNRAS.477....2J}). This will open a new window of \hi\ mass function measurements outside the local Universe.
\end{itemize}

To summarize, we introduce a new \hi\ halo model as an analytical framework to model the \hi\ power spectrum. Using it, we can analyse the \hi\ power spectrum to break the degeneracy between \hi\ density and \hi\ bias, which is essential in order to use the \hi\ intensity mapping power spectrum for cosmology. The measurement of \hi\ shot noise can be used to constrain the \hi\ galaxy distribution at various redshifts. 
We will be working towards including more complex issues such as foreground removal and redshift-space distortions, and to applying our framework to real data in the future.

\section*{Acknowledgements}
Besides aforementioned ones, this work uses open-source packages NumPy \citep{5725236}, SciPy \citep{2020SciPy-NMeth}, IPython \citep{4160251}, {\tt ASTROPY} \citep{astropy:2013,astropy:2018}, {\tt MATPLOTLIB} \citep{Hunter:2007} and {\tt CORNER} \citep{corner}. ZC acknowledges support from the Overseas Research Scholar Awards from
the School of Physics and Astronomy, The University of
Manchester.

\section*{Data Availability}
The computational tool used in this paper is available at \url{https://github.com/steven-murray/halomod}. The simulation data underlying this paper will be shared on reasonable request to the corresponding author.

\bibliographystyle{mnras}
\bibliography{references}

\appendix
\section{Analytical Expression of \althi\ Shot Noise in the Halo Model}
\label{app:sn}
We present the derivation of Eq. (\ref{eq:PSN}) here. We start with Eq. (\ref{eq:PSNsim}):
\begin{equation}
    P_{\rm SN} = V\frac{\Big\langle\sum\limits_i \big(M_{\rm \hi}^{i}\big)^2\Big\rangle}{\Big(\big\langle \sum\limits_i M_{\rm \hi}^{i}\big\rangle\Big)^2}
\end{equation}
where $i$ run over all \hi\ sources.

To put it into the context of halos, we rewrite it as
\begin{equation}
    P_{\rm SN} = \frac{1}{V\Bar{\rho}_{\rm \hi}^2}\Big\langle\sum_i^{N_{\rm h}}\sum_j^{N_{\rm g}^i}
    \big(M_{\rm \hi}^{ij}\big)^2\Big\rangle
\end{equation}
where $N_{\rm h}$ is the number of halos with a number of $N_{\rm g}^i$ galaxies in the $i^{\rm th}$ halo and $M_{\rm \hi}^{ij}$ is the mass of \hi\ inside the $j^{\rm th}$ galaxy of the $i^{\rm th}$ halo.

Usually central and satellite components are modelled separately. Therefore, we rewrite the above equation as:
\begin{equation}
    P_{\rm SN} = \frac{1}{V\Bar{\rho}_{\rm \hi}^2}\Big\langle\sum_i^{N_{\rm h}}\sum_{k}^{\rm cen,sat}\sum_j^{N_{\rm g}^{i,k}}
    \big(M_{\rm \hi}^{ij,k}\big)^2\Big\rangle.
\end{equation}

$M_{\rm \hi}^{ij,k}$ follows an unknown random distribution with a mean $\big\langle M_{\rm \hi,field}^k (M_{\rm h}^i) \big\rangle$. Therefore, we model it as
\begin{equation}
   M_{\rm \hi}^{ij,k} = \big\langle M_{\rm \hi,field}^k (M_{\rm h}^i) \big\rangle\Big(1+\sigma_{\rm \hi}^k(M_{\rm h}^i) x_j\Big) 
\label{eq:mijk}
\end{equation}
where $x_j$ follows an unknown random distribution with $\langle x_j\rangle = 0$ and $\langle x_j^2\rangle = 1$, such that
\begin{align}
    \big\langle M_{\rm \hi}^{ij,k}\big\rangle &= \big\langle M_{\rm \hi,field}^k (M_{\rm h}^i) \big\rangle,\\
    {\rm Std}\bigg(M_{\rm \hi}^{ij,k}\bigg) &= \big\langle M_{\rm \hi,field}^k (M_{\rm h}^i) \big\rangle \sigma_{\rm \hi}^k(M_{\rm h}^i).
\label{eq:std}
\end{align}
Each halo has a number of \hi\ galaxies within. We model its randomness:
\begin{equation}
    N_{\rm g}^{i,k} = \big\langle N_{\rm g}^k(M_{\rm h}^i) \big\rangle\big
    (1+\sigma_{\rm g}^k(M_{\rm h}^i) x_i\big).
\label{eq:nik}
\end{equation}
$x_i$ follows another unknown random distribution with zero mean and a standard deviation of 1 such that
\begin{equation}
    {\rm Std}\bigg(N_{\rm g}^{i,k}\bigg) = \big\langle N_{\rm g}^{i,k}\big\rangle \sigma_{\rm g}^k(M_{\rm h}^i).
\end{equation}
The number density of galaxies and the \hi\ as within is related to the \hi\ HOD:
\begin{equation}
    \Big\langle N_{\rm g}^{i,k} M_{\rm \hi}^{ij,k} \Big\rangle = \Big\langle M_{\rm \hi}^{k}\big(M_{\rm h}^i\big)\Big\rangle.
\end{equation}
Substituting Eqs. (\ref{eq:mijk}) and (\ref{eq:nik}) into the above equation we have
\begin{equation}
\begin{split}
   \Big\langle M_{\rm \hi}^{k}\big(M_{\rm h}^i\big)\Big\rangle&=
   \bigg\langle \big\langle N_{\rm g}^k(M_{\rm h}^i) \big\rangle
    \big\langle M_{\rm \hi,field}^k (M_{\rm h}^i) \big\rangle
    \big(1+\sigma_{\rm g}^k(M_{\rm h}^i)x_i\big)\\
    &\times\big(1+\sigma_{\rm \hi}^k(M_{\rm h}^i)x_j\big)
    \bigg\rangle\\
    = \big\langle N_{\rm g}^k(M_{\rm h}^i) \big\rangle&
    \big\langle M_{\rm \hi,field}^k (M_{\rm h}^i) 
    \big\rangle\Big(1+\sigma_{\rm g}^k(M_{\rm h}^i)\sigma_{\rm \hi}^k(M_{\rm h}^i) \big\langle x_ix_j \big\rangle\Big).
\end{split}
\end{equation}
We can further simplify as
\begin{equation}
    \big\langle M_{\rm \hi,field}^k (M_{\rm h}^i) \big\rangle = \frac{\big\langle M_{\rm \hi}^k(M_{\rm h}^i)\big\rangle}{\big\langle N_{\rm g}^k(M_{\rm h}^i)\big\rangle}\bigg/\Big(1+\sigma_{\rm g}^k(M_{\rm h}^i)\sigma_{\rm \hi}^k(M_{\rm h}^i) \big\langle x_ix_j \big\rangle\Big).
\label{eq:hodfield}
\end{equation}
With this definition we can rearrange the shot noise as
\begin{equation}
\begin{split}
    P_{\rm SN} =& \frac{1}{V\Bar{\rho}_{\rm \hi}^2}\bigg\langle\sum_i^{N_{\rm h}}\sum_{k}^{\rm cen,sat}\sum_j^{N_{\rm g}^{i,k}}\Big[\big\langle M_{\rm \hi,field}^k (m) \big\rangle\big(1+\sigma_{\rm \hi}^k(M_{\rm h}^i) x_j\big)\Big]^2\bigg\rangle\\
    =& \frac{1}{\Bar{\rho}_{\rm \hi}^2} \bigg\langle\sum_i^{N_{\rm h}} \int {\rm d}m\; \frac{\delta_D(m-M_{\rm h}^i)}{V}\\&\times \sum_{k}^{\rm cen,sat}\sum_j^{N_{\rm g}^k(m)}\Big[\big\langle M_{\rm \hi,field}^k (m) \big\rangle\big(1+\sigma_{\rm \hi}^k(m) x_j\big)\Big]^2\bigg\rangle.
\end{split}
\label{eq:psnapp}
\end{equation}
Note that the abundance of halo is uncorrelated with the distribution of \hi\ inside it and therefore we can extract
\begin{equation}
    n(m)\equiv \big\langle \sum_i^{N_{\rm h}}\frac{\delta_D(m-M_{\rm h}^i)}{V}\big\rangle.
\end{equation}
Now we rewrite Eq. (\ref{eq:psnapp}) as
\begin{equation}
\begin{split}
    P_{\rm SN} 
    =& \frac{1}{\Bar{\rho}_{\rm \hi}^2}\sum_{k}^{\rm cen,sat}
    \int {\rm d}m\; n(m) \big\langle M_{\rm \hi,field}^k(m)\big\rangle^2\\
    &\times\bigg\langle \sum_j^{N_{\rm g}^k(m)}\big(1+\sigma_{\rm \hi}^k(m) x_j\big)^2 \bigg\rangle\\
    =& \frac{1}{\Bar{\rho}_{\rm \hi}^2}\sum_{k}^{\rm cen,sat}
    \int {\rm d}m\; n(m) \big\langle M_{\rm \hi,field}^k(m)\big\rangle^2 \\&\times\bigg\langle \big\langle N_{\rm g}^k(m) \big\rangle
    \times\big(1+\sigma_{\rm g}^k(m) x_i\big) \big(1+\sigma_{\rm \hi}^k(m) x_j\big)^2 \bigg\rangle.
\end{split}
\end{equation}
This results in
\begin{equation}
\begin{split}
    P_{\rm SN} 
    = &\frac{1}{\Bar{\rho}_{\rm \hi}^2}\sum_{k}^{\rm cen,sat}
    \int {\rm d}m\; n(m) \big\langle M_{\rm \hi,field}^k(m)\big\rangle^2
    \big\langle N_{\rm g}^k(m) \big\rangle \\ 
    &\times \bigg(1+2\sigma_{\rm g}^k(m)\sigma_{\rm \hi}^k(m)\big\langle x_i x_j \big\rangle + \big(\sigma_{\rm \hi}^k(m)\big)^2 \\&+ \sigma_{\rm g}^k(m)\big(\sigma_{\rm \hi}^k(m)\big)^2\big\langle x_i x_j^2 \big\rangle\bigg)\\
    = &\frac{1}{\Bar{\rho}_{\rm \hi}^2}\sum_{k}^{\rm cen,sat}
    \int {\rm d}m\; n(m) \Big( \frac{\big\langle M_{\rm \hi}^k(M_{\rm h}^i)\big\rangle}{\big\langle N_{\rm g}^k(M_{\rm h}^i)\big\rangle}\Big)^2
    \big\langle N_{\rm g}^k(m) \big\rangle \\ 
    &\times \bigg(1+2\sigma_{\rm g}^k(m)\sigma_{\rm \hi}^k(m)\big\langle x_i x_j \big\rangle + \big(\sigma_{\rm \hi}^k(m)\big)^2 \\
    &+ \sigma_{\rm g}^k(m)\big(\sigma_{\rm \hi}^k(m)\big)^2\big\langle x_i x_j^2 \big\rangle\bigg)
    \bigg/\Big(1+\sigma_{\rm g}^k(M_{\rm h}^i)\sigma_{\rm \hi}^k(M_{\rm h}^i) \big\langle x_ix_j \big\rangle\Big)^2.
\end{split}
\label{eq:PSNfull}
\end{equation}
If we assume that the scatter of galaxy HOD and scatter of \hi\ within it do not correlate, i.e. $\langle x_ix_j \rangle =0$ and $\langle x_ix_j^2 \rangle =0$, the equation simplifies to
\begin{equation}
\begin{split}
    P_{\rm SN} = \frac{1}{\Bar{\rho}_{\rm \hi}^2}\sum_{k}^{\rm cen,sat}
    \int &{\rm d}m\; n(m) \big\langle M_{\rm \hi}^k(m)\big\rangle^2 \big\langle N_{\rm g}^k(m)\big\rangle^{-1}\\
    &\times\Big(1+\big(\sigma_{\rm \hi}^k(m)\big)^2 \Big). 
\end{split}
\end{equation}
\bsp
\label{lastpage}
\end{document}